\documentclass[acmsmall]{acmart}

\usepackage{tabularx}
\usepackage{adjustbox}
\usepackage{tcolorbox}
\usepackage{subcaption}
\usepackage{pifont}
\usepackage{enumitem}
\usepackage{makecell}
\usepackage[nameinlink]{cleveref}
\usepackage{multirow}
\usepackage{threeparttable}
\usepackage{soul}
\usepackage{colortbl}

\newboolean{showcomments}
\setboolean{showcomments}{false}
\ifthenelse{\boolean{showcomments}}
{\newcommand{\mynote}[2]{
\fbox{\bfseries\sffamily\scriptsize#1}
{\small$\blacktriangleright$\textsf{\emph{#2}}$\blacktriangleleft$}}}
{ \newcommand{\mynote}[2]{}}

\def\BibTeX{{\rm B\kern-.05em{\sc i\kern-.025em b}\kern-.08em
    T\kern-.1667em\lower.7ex\hbox{E}\kern-.125emX}}

\newcommand{\find}[1]{
\begin{tcolorbox}[leftrule=0.4mm,rightrule=0mm,toprule=0mm,bottomrule=0mm,left=0.0pt,right=0.0pt,top=0pt,bottom=0pt]
\em #1
\end{tcolorbox}
}

\setcopyright{acmlicensed}
\acmDOI{10.1145/3715764}
\acmYear{2025}
\acmJournal{PACMSE}
\acmVolume{2}
\acmNumber{FSE}
\acmArticle{FSE047}
\acmMonth{7}
\received{2024-09-13}
\received[accepted]{2025-01-14}

\begin{document}

\title{The Struggles of LLMs in Cross-Lingual Code Clone Detection}

\author{Micheline Bénédicte Moumoula}
\orcid{0009-0005-1206-2270}
\authornote{Corresponding author (micheline.moumoula@uni.lu)}
\affiliation{%
  \institution{University of Luxembourg }
  \city{Luxembourg}
  \country{Luxembourg}
}
\affiliation{%
  \institution{Centre d'Excellence Interdisciplinaire en Intelligence Artificielle pour le Développement (CITADEL)}
  \city{Ouagadougou}
  \country{Burkina Faso}
}
\email{micheline.moumoula@uni.lu}

\author{Abdoul Kader Kaboré}
\orcid{0000-0002-3151-9433}
\affiliation{%
  \institution{University of Luxembourg}
  \city{Luxembourg}
  \country{Luxembourg}
}
\email{abdoulkader.kabore@uni.lu}

\author{Jacques Klein}
\orcid{0000-0003-4052-475X}
\affiliation{%
  \institution{University of Luxembourg}
  \city{Luxembourg}
  \country{Luxembourg}
}
\email{jacques.klein@uni.lu}

\author{Tegawendé F. Bissyandé}
\orcid{0000-0001-7270-9869}
\affiliation{%
  \institution{University of Luxembourg}
  \city{Luxembourg}
  \country{Luxembourg}
}
\email{tegawende.bissyande@uni.lu}


\begin{abstract}

With the involvement of multiple programming languages in modern software development, cross-lingual code clone detection has gained traction within the software engineering community. Numerous studies have explored this topic, proposing various promising approaches.
Inspired by the significant advances in machine learning in recent years, particularly Large Language Models (LLMs), which have demonstrated their ability to tackle various tasks, this paper revisits cross-lingual code clone detection. 
We evaluate the performance of five (05) LLMs and, eight prompts (08) for the identification of cross-lingual code clones. Additionally, we compare these results against two baseline methods. Finally, we evaluate a pre-trained embedding model to assess the effectiveness of the generated representations for classifying clone and non-clone pairs. The studies involving LLMs and Embedding models are evaluated using two widely used cross-lingual datasets, XLCoST and CodeNet.

Our results show that LLMs can achieve high F1 scores, up to 0.99, for straightforward programming examples. However, they not only perform less well on programs associated with complex programming challenges but also do not necessarily understand the meaning of ``code clones'' in a cross-lingual setting. 
We show that embedding models used to represent code fragments from different programming languages in the same representation space enable the training of a basic classifier that outperforms all LLMs by $\sim$1 and $\sim$20 percentage points on the XLCoST and CodeNet datasets, respectively. This finding suggests that, despite the apparent capabilities of LLMs, embeddings provided by embedding models offer suitable representations to achieve state-of-the-art performance in cross-lingual code clone detection.

\end{abstract}


\maketitle
\section{Introduction} 
\label{sec:introduction}

Copy-pasting a code fragment with or without change is common practice in software development. It leads to what is commonly referred to as code clones~\cite{roy_survey_nodate}. 
While clones provide some benefits (e.g., faster prototyping, reuse of proven code), they are also known to contribute to reducing the quality of the code, making maintenance costly, and being a source of bugs~\cite{juergens_code_2009, roy_survey_nodate}. 
Research shows that we can find between 5\% to 23\% of clones in a software system~\cite{white_deep_2016, koschke_survey_nodate}. 
The literature enumerates four types of clones~\cite{keller_what_2022}. 
Types 1, 2, and 3 clones are generally classified as syntactic clones, and type 4 is referred to as semantic clones. 
Syntactic clones occur when there are textual similarities, while semantic clones represent functional similarity. 
Type 4 is the most complicated clone and the hardest to detect \cite{dou_towards_2023}.

Code clone retrieval is generally studied in the context of program fragments written with the same programming language. 
However, in modern software development, the multiplicity of software components involves various programming languages used, each for each efficiency in addressing specific programming challenges~\cite{kochhar_large_2016}. 
Developers may for example intentionally introduce clones for building identical systems for different platforms.
Since the development of those systems is, for the most part, collaborative,  if an expert in a certain type of programming language changes one part of the software for a given language, the exact functional change must be carried over to the other variants. 
This is more resource and time-consuming than modifying a software system based on a single language, as it requires prior knowledge of the system architecture and an understanding of the code changes performed by the first developer \cite{nafi_clcdsa_2019}.
To manage cross-lingual systems in an easy, time-effective, and cost-effective way, developers need an automatic system that can detect clones of various languages at the same time\cite{nafi_clcdsa_2019}. 

The literature includes several approaches and tools for code clone detection. For instance, CCFinder ~\cite{kamiya_ccfinder_2002}, Deckard ~\cite{jiang_deckard_2007}, NiCad ~\cite{roy_nicad_2008}, SourcererCC~\cite{sajnani_sourcerercc_2016}, NIL~\cite{nakagawa_nil_2021}, and OREO~\cite{saini_oreo_2018} have been developed for type 1, 2, and 3 clones detection within the same programming language. 
For semantic clones within the same language, RtvNN ~\cite{white_deep_2016}, CDLH ~\cite{wei_supervised_2017}, DeepSim~\cite{zhao_deepsim_2018}, and FA-AST ~\cite{wang_detecting_2020} have been proposed.

Concerning cross-lingual clones detection, tools such as LICCA~\cite{vislavski_licca_2018}, CLCMiner \cite{cheng_clcminer_2017}, CLCDSA~\cite{nafi_clcdsa_2019}, OneSpace~\cite{el2024onespace}, C4~\cite{tao_c4_2022}, TCCCD~\cite{fang_tcccd_2023} , CCT-Code~\cite{sorokin_cct-code_2023}, AdaCCD ~\cite{du_adaccd_2024} and ~\cite{khajezade_investigating_2024} have been introduced. 
They rely on deep learning techniques to capture the syntactic and semantic relationships between different parts of the source code.
CLCDSA employed on-the-fly and deep neural network-based approaches using Abstract Syntax Trees (ASTs), a fundamental data structure in computer science for representing code structure, to extract features for cross-lingual code clone detection. 
On the fly approach measures cosine similarity between the two matrices generated by extracting values of the features from source code fragments while the deep neural network-based approach learns automatically the values of features from data and detects cross-lingual clones. 
OneSpace introduces a novel approach to cross-language code clone detection by mapping programming languages into a shared embedding space using both code and API data. It employs a Siamese network to evaluate the similarity between embedded programs and explores an alternative method using Word Mover's Distance for efficiency. The approach also examines the influence of factors such as code complexity and training data size on its performance.
TCCCD employed the UniXcoder ~\cite{guo_unixcoder_2022} pre-trained model to map code clones from different languages into vector space and fine-tuned the model using triplet learning. 
Sorokin et al. proposed an approach using cross-consistency training (CCT) for language models training on source code in different programming languages and presented the CCT-LM model, initialized with GraphCodeBERT and fine-tuned with CCT to retrieve semantically similar codes given a code as the query. 
AdaCCD leverages language-agnostic representations from pre-training programming language models (GraphCodeBERT and CodeBERT) and contrastive learning for cross-lingual adaptation to detect code clones.

The rise in recent years of Large Language Models (LLMs) has led to the introduction of new studies in various domains, including software engineering.
As regards code clone detection, the researcher also investigated the use of LLMs.

A representative work on this topic is proposed by Dou et al.~\cite{dou_towards_2023}, who proposed an approach with diverse prompt engineering techniques to assess LLMs' ability to detect clones in several programming languages. 
The experimental results showed that GPT-3.5-turbo and GPT-4 \footnote{https://openai.com/} excel in detecting complex semantic clones within the same programming language, surpassing existing methods. However, this study only focuses on code clone detection in the same language. 

In this paper, we investigate Large Language Models (LLMs), and Embedding Models' (EM) capabilities for cross-lingual code clone detection.
LLMs performances being influenced by prompt engineering~\cite{lin2024write,baiexploring,yin2024should}, we first design various different prompts that we applied on five LLMs (Falcon-7B-Instruct ~\cite{falcon40b}, LLAMA2-Chat-7B~\cite{touvron_llama_2023}, Starchat-$\beta$ ~\cite{Tunstall2023starchat-alpha}, StarCoder2-15b-Instruct~\cite{StarCoder2-Instruct} and GPT-3.5-Turbo).
The design prompts include two simple prompts: 
one that asks the LLMs to directly provide "yes/no" answers if the two code snippets are clones or not, and an improved version that includes instructions asking the LLMs to take into account the whole structure and logic of the two code snippets and conclude if they perform a similar task.
In addition to those simple prompts, we designed one-step and multi-step Chain of Thought (CoT)~\cite{dou_towards_2023} prompts to evaluate their impact on LLMs' performances. With these prompts, we want the LLMs to follow step-by-step reasoning to perform the task by taking into account the code similarity, the similar lines, the thinking process, and the code description analysis.

Our second study for cross-lingual code clone detection relies on the use of embedding models.
We selected three pre-trained models: GraphCodeBERT ~\cite{guo2020graphcodebert} and UniXcoder ~\cite{guo2022unixcoder}, two widely used open-source embedding models for both natural language (e.g., descriptions, comments) and programming languages (source code), and the `Text-embedding-3-large' provided by OpenAI.

For GraphCodeBERT and UniXcoder models, which are also used as baselines, similar to prior work~\cite{fang_tcccd_2023, khajezade_investigating_2024} we directly leverage their ability to be modified for downstream tasks (in our case, cross-lingual code clone detection).

For the `Text-embedding-3-large' model, given two code snippets, we directly provide them as inputs to the embedding model, which outputs two vectors.
To detect whether the provided code snippets are clones or not, we can directly perform a similarity measure using the two vectors and cosine similarity~\cite{pedregosa2011scikit}. 
This requires the definition of a threshold. 
In our experiments, we investigate several threshold values. 
With the embedding vectors, we also conducted additional experiments for clone detection by training custom binary classifiers.

We perform evaluations using both the XLCoST~\cite{zhu_xlcost_2022} and CodeNet~\cite{puri_codenet_2021} datasets. 
We identified those datasets with respect to the large number of samples included and the various programming covered by covered languages.
While the datasets have been used in the literature for many cross-lingual related tasks~\cite{muennighoff2022crosslingual, gu2024cruxeval, khajezade_investigating_2024, li2023zc}, they are not directly usable as benchmarks for code clone detection: \ding{172} clone pairs must be established, and non-clone pairs must be carefully constructed. We undertake this effort and contribute to the community with a benchmark that includes a balanced set of 6\,000 positive pairs (i.e., clones) and 6\,000 negative pairs (i.e., non-clones).

Our experimental results show that GPT-3.5-Turbo achieved an overall F1 score {of 0.99 for the XLCoST and 0.82} for CodeNet. 
It outperforms LLAMA2-Chat-7B, Starchat-$\beta$, StarCoder2-Instruct~\cite{StarCoder2-Instruct}  and Falcon-7B-Instruct that achieve respectively overall F1 scores of 0.82, 0.615, 0.595, and 0.48  for the XLCoST dataset and 0.58, 0.52, 0.47, and 0.43 for the CodeNet dataset.
Among all the designed prompts, the ``improved simple prompt'' is the one that enables all LLMs to achieve their best performance in the task of code clone detection. This underlines that LLMs' understanding of the notion of clones is narrow but that, once provided with definition guidance, they are relatively able to reason across programming languages.

Regarding the embedding model, we first note that the cosine similarity derived from code snippets' vectors can be directly used to identify code clones.
Indeed, with cross-lingual pairs, the embedding model produces close vectors when the code snippets are clones and distant vectors when they are non-clones.
This is observed in $\sim$74\% of the pairs with a cosine similarity threshold of 50\%.
We further leverage the vector pairs to train several classifiers for cross-lingual code clone detection.
Among those classifiers, the Support Vector Machine classifier (SVM) using a polynomial kernel achieves F1-scores of 1 and 0.986, respectively, for XLCoST and CodeNet.
In sum, all our evaluations reveal that the embedding model outperforms all LLMs, even though the latter yields satisfactory results when combined with  CoT-based prompts.

This research addresses the challenging problem of cross-language clone detection, which requires a deeper semantic understanding of code snippets across diverse syntaxes, structures, and programming styles. By leveraging the capabilities of large language models (LLMs), we explore their potential and limitations in this complex domain and propose strategies, including enhanced prompt designs, to improve their effectiveness.
We introduce an approach that combines binary classifiers with code embeddings to effectively capture semantic similarities in cross-language scenarios. This method offers a practical solution for analyzing multilingual codebases, addressing critical challenges in areas such as multilingual code refactoring, security analysis, and software maintenance.
By focusing on the application of LLMs to cross-language clone detection, this research contributes to an underexplored area with significant implications for managing large and diverse software ecosystems.
This work makes several key contributions to the field of cross-lingual code clone detection:
\begin{itemize}[leftmargin=*]
  \item We conduct a large-scale investigation of LLMs' capabilities for cross-lingual code clone detection. Our study considers Java paired against 10 programming languages, making it the largest study applying various LLMs to various cross-lingual datasets. Furthermore, compared to existing literature, we assess the influence of different prompt engineering techniques. 
  
  \item  We provide insightful findings on the performance of LLMs in code clone detection. Notably, we show that the similarity between two programming languages influences the ability of LLMs to detect code clones across them, mainly when the prompt is simple. If the prompt instructs the LLM to focus on logic/reasoning, programming language differences have substantially less influence. Eventually, we discuss the generalizability and overall effectiveness of LLMs for the task of cross-lingual code clone detection.
  
  \item We discuss the performance of LLMs by comparing them against two baselines and the classical approach of machine learning classification based on learned representations of code. Experimental results even hint that LLMs may not even ``understand" the meaning of "clones"  in the context of clones.  
\end{itemize}

The remainder of this paper is organized as follows: \Cref{sec:background} presents background. \Cref{sec:experimental_setup} introduces the experimental setup. \Cref{sec:evaluation} evaluates our analysis. \Cref{sec:related-work} discusses related work. \Cref{sec:conclusion} concludes.
\section{Background}
\label{sec:background}

This section presents a short introduction to the concepts of code clone and LLMs.

\subsection{Code Clones}

A pair of code clones represents two code fragments that are similar, by some given definition of similarity~\cite{roy2009comparison}. We can enumerate four types of clones : 
\begin{itemize}[leftmargin=*]
    \item Type-1 or identical code fragments represent the same code except for white space, comments, and layout.
    \item Type-2 or lexical code snippets represent identical clone pairs except for differences in variables or function names with Type-1 clone differences.
    \item Type-3 or syntactically represent similar code fragments that differ at the statement level. The code fragments differ in some lines with removed or added of some lines in addition to type-2 clone differences. 
    \item Type-4 or semantic code clone represents code snippets that perform the same functionality but the implementation is different. In global they are syntactically dissimilar.
    \item Cross lingual clone is a semantic clone when code fragments are written in different programming languages. Our work is focused on semantic code clone detection for cross-lingual code snippets.
\end{itemize}

\subsection{Large Language Models}
LLMs are a type of artificial intelligence (AI) trained on massive datasets of text and code. This training enables them to perform a variety of natural language processing (NLP) tasks, including text summarization, text completion, machine translation, and even creative text generation. While LLMs are still under development, they have demonstrated the ability to generate human-quality responses to prompts.

Prompts are instructions or queries provided to an LLM to guide its output towards a specific task. Prompt engineering is a crucial aspect of LLM performance and refers to the optimization of prompts to achieve the most accurate and desirable outputs~\cite{sahoo2024systematic}. Chain-of-thought prompting is a specific technique that utilizes a sequence of connected instructions to guide the LLM's reasoning process step-by-step, resulting in demonstrably more accurate solutions for complex problems

\section{Experimental Setup}
\label{sec:experimental_setup}

First, we articulate the research questions that we aim to address. Second, we introduce the models that are leveraged in this study before presenting the datasets and describing the metrics. 
Finally, we introduce our experimental methodology on applying LLMs and embedding models for cross-lingual code clone detection. 

\subsection{Research Questions} \label{rqs}
We conduct our study around five research questions. 
The first four research questions are related to evaluating LLMs' performances in detecting cross-lingual code clones.
In the last research question, we also evaluated the effectiveness of an embedding model and compared its performance against LLMs.

\begin{itemize}[leftmargin=*]

    \item \textbf{RQ1}:  {\em To what extent does prompt engineering improve the effectiveness of LLMs in detecting cross-lingual code clones?} With this research question, we provide empirical results on the effectiveness of LLMs for the cross-lingual code clone detection task. The experiment comprehensively considers various prompting strategies to provide insights into the impact of prompt designs across several LLMs.

    \item \textbf{RQ2}: {\em To what extent do LLMs actually understand the task of code clone detection in a cross-lingual setting?} Code clone detection is about comparing the similarity of code fragments. In a cross-lingual setting, an LLM must make abstractions to achieve similarity comparisons. With this research question, we investigate how LLMs define ``clones'' and whether they can be guided towards making the necessary abstractions for the task of code clone detection in a multi-lingual setting.

    \item \textbf{RQ3}: {\em How does the similarity between programming languages influence the performance of LLMs in cross-lingual code clone detection?} Because programming languages may or may not share similarities in terms of syntax, we hypothesize that LLMs performance may be affected by different configurations of the cross-lingual clone detection tasks. 
    
    \item \textbf{RQ4}: {\em Are LLMs outperforming classification models for cross-lingual code clone detection?} Before the advent of LLMs, state-of-the-art approaches for code clone detection were employing machine learning-based classification, notably using deep representation learning. We thus compare LLMs against baseline classifiers (e.g. with SVM and kNN) based on Embedding models. The objective is to shed lights into the power of representations vs reasoning complexity in LLMs.

\end{itemize}

\subsection{Models}
\label{lab:models}
We evaluated five LLMs and one embedding model for cross-lingual code clone detection. The selection process prioritized models fine-tuned for programming languages and code-related tasks, as well as those offering reproducibility and accessibility. From OpenAI, we included the popular GPT-3.5-Turbo model due to its prevalence in software engineering research and its balance of capability and cost-effectiveness, making it suitable for large-scale evaluations. 
To explore open-source options, we include Falcon-7B-Instruct, Starchat-$\beta$, and LLAMA2-Chat-7B. These models are trained on multiple programming languages and some have been employed in prior studies for code clone detection within the same programming language. Among these, StarCoder2-15b-Instruct was selected for its robust fine-tuning on programming languages, making it particularly well-suited for our task.
Additionally, we incorporated the Text-embedding-3-large model, an embedding model from OpenAI's web API, to investigate the effectiveness of pre-trained embeddings for this task. By including this model, we extended our analysis beyond LLMs to assess the broader potential of embedding-based approaches. we provide detailed descriptions of each model in the following section.
\begin{itemize}[leftmargin=*]
    \item GPT-3.5-Turbo is an LLM developed by OpenAI, known for its strong performance in various language tasks. It excels at understanding and generating text in a conversational context, producing coherent and relevant responses based on user input. Notably, the model can retain information short-term, facilitating meaningful dialogue. Trained on a massive dataset of internet text, GPT-3.5-Turbo possesses a broad knowledge base and can leverage this information for diverse language processing applications.

    \item The Falcon-7B-Instruct model is a large language model (LLM) with 7 billion parameters. It is based on the Falcon-7B architecture and has been fine-tuned on a combined dataset of conversational and instructional text. This fine-tuning process optimizes the model for performing tasks typically associated with virtual assistants.
    
    \item The LLaMA2-Chat-7B model is a large language model (LLM) with 7 billion parameters. It is trained on a large corpus of text data, including conversational elements such as chat logs and social media posts. This training allows the model to acquire the patterns and structures of natural language dialogue, enabling it to generate coherent and contextually relevant responses to user prompts.
    
    \item StarChat-$\beta$ is an instruction-tuned large language model (LLM) designed to act as a helpful coding assistant. The model is based on StarCoderPlus ~\cite{li_starcoder_2023}, a 15.5-billion parameter LLM trained on English and over 80 programming languages. StarChat-$\beta$ is further fine-tuned on a specifically designed, "uncensored" variant of the open-assistant guanaco dataset. This fine-tuning process strengthens the model's ability to understand and respond to human instructions within the context of coding tasks.

    \item StarCoder2-15b-Instruct ~\cite{StarCoder2-Instruct} is the first code LLM trained entirely through a self-aligned process, eliminating the need for human annotations or distilled data from proprietary models. Our open-source pipeline leverages StarCoder2-15B ~\cite{starcoder2} to generate a substantial dataset of instruction-response pairs. These pairs are then used to iteratively refine the model's capabilities, ensuring a fully transparent and permissive training process.

    \item Text-embedding-3-large is a text embedding model created by OpenAI. It is designed for tasks such as search, clustering, recommendations, anomaly detection, diversity measurement, classification text similarity evaluation, and code search. The model generates numerical representations (embeddings) of text or code inputs. Embeddings are low-dimensional, dense vector representations that capture the semantic relationships between words or code elements. This allows computers to more efficiently process and understand the meaning of text and code.
    
\end{itemize}

\noindent
\textbf{Baselines.}
We compare the results of the LLMs against two baseline models, chosen for their robust performance in previous research, to strengthen our evaluation.

\begin{itemize}[leftmargin=*]
    \item The GraphCodeBERT ~\cite{guo_graphcodebert_2021}  is a transformer-based model that prioritizes data flow over traditional syntactic structures, such as abstract syntax trees (AST), in its pre-training phase. By leveraging structure-aware approaches, it enhances code representation, making it particularly effective for tasks like code clone detection, translation, refinement, and search. This focus on the intrinsic flow of data within code enables GraphCodeBERT to capture deeper semantic relationships, leading to improved performance across various code understanding tasks.
    \item The Unixcoder  ~\cite{guo_unixcoder_2022} is a cross-modal pre-trained model for programming languages that improves upon encoder-decoder frameworks by using mask attention matrices with prefix adapters for fine-grained control. It leverages data such as abstract syntax trees (ASTs) and code comments to enhance source code representation. Pre-trained on nine programming languages, including  Java, Ruby, Python, PHP, Javascript, Go, C, C++, and C\#, UniXcoder offers enhanced capabilities in representing cloned code, particularly in C++ and C\#.
\end{itemize}

\subsection{Datasets}
\label{lab:datasets}
We selected two cross-lingual benchmarks to ensure a robust and comprehensive evaluation across all considered programming languages: XLCoST \cite{zhu_xlcost_2022} and CodeNet \cite{puri_codenet_2021}. 
\begin{itemize}[leftmargin=*]
    \item The \textbf{XLCoST} dataset~\cite{zhu_xlcost_2022} is collected from GeeksForGeeks\footnote{https://www.geeksforgeeks.org/}, a popular online platform for computer science education and skill development. Notably, it hosts coding challenges for programmers. The dataset includes samples from seven different programming languages (C, C++, C\#, Java, JavaScript, PHP, Python ). Since the dataset includes only true pairs of cross-lingual code clones, we must undertake to create negative clone pairs.

    \item The \textbf{CodeNet} dataset~\cite{puri_codenet_2021} comprises a large collection of code samples with extensive metadata across 4000 coding problems and over 50 programming languages (although C++, C, Python, and Java are the dominant languages). The code samples are annotated with a rich set of information, such as the problem ID and the status, which indicates acceptance or error types. The data is collected from AIZU Online Judge \footnote{https://onlinejudge.u-aizu.ac.jp/} and AtCoder \footnote{https://atcoder.jp/}. Based on the problem ID, we can construct both, positive and negative clone pairs

\end{itemize}

\noindent
\textbf{Construction of sample pairs.}
To ensure the highest quality samples for our experiments, we implemented a rigorous selection process for both non-clone and clone pairs.
Both raw datasets include samples with a problem ID, description, and code snippets in various programming languages. This structure allows us to easily retrieve positive pairs, defined as “two code snippets in different programming languages addressing the same programming problem”. However, it is challenging to build a ground truth set of negative pairs. We postulate that a pair can be negative for two main reasons: (i) the problems addressed by the two code snippets are different; (ii) for the same problem two code snippets actually differ semantically (e.g., one is incorrect). For constructing our dataset, we only select code validated by developers. This means that negative pairs are only those where the problems are different. We therefore propose to use sentence embeddings to compare problem descriptions and build the dataset of negative pairs. Here is the breakdown:
\begin{itemize}[leftmargin=*]
    \item[\ding{182}] For each problem description, we generate the corresponding sentence vector using a state-of-the-art sentence vector generation model (all-mpnet-base-v2 \footnote{https://huggingface.co/sentence-transformers/all-mpnet-base-v2}). This model has been widely validated for its effectiveness in capturing semantic meaning.

    \item[\ding{183}] To group similar problems, we applied the DBSCAN clustering algorithm to the sentence vectors. Within each cluster, we selected the most complicated problem based on a pre-computed complexity metric. We identify the furthest problem (by vector distance) as a candidate for a negative pair (considering all the problems in the dataset). 
    
    \item[\ding{184}] \textbf{Negative set selection}: We ensure that there are no duplicates or over-representation of specific problems and we prioritize pairing the most complex code.
    
    \item[\ding{185}] \textbf{Positive set selection}: The remaining problems were used to construct the positive set, with the most complex code selected to represent the clone pairs.
    
    Before the previously described process, we calculated the cyclomatic complexity of each problem and code sample provided by the two datasets selecting the most complex instances to ensure a representative and challenging dataset.    
\end{itemize}
 For our experiments, we used subsets of 3000 clone pairs and 3000 non-clone pairs from each dataset, with Java selected as the primary language cloned alongside four and ten programming languages, respectively, for the XLCoST and, the CodeNet dataset with the same ratio for each programming language.

\textbf{For the Baselines}, We employ a similar selection process. However, we restricted our dataset to CodeNet, given its broader coverage of programming languages. From CodeNet, we selected only the programming languages that were compatible with the baseline models. Additionally, we filtered the dataset to include only pairs that met the input size requirements of the baseline models.

\subsection{Metrics}
Code clone detection can be considered as a binary classification problem. Therefore, we rely on classical metrics for classification evaluation, notably Precision, Recall, and F1 score. 
Precision represents the proportion of true, correct positive predictions to all of the positive predictions (TP/(TP+FP)). 
The recall represents the proportion of true correct positive prediction to all of the true positive samples (TP/(TP+FN)). 
F1 score is defined as the harmonic mean of precision and recall.

\subsection{Methodology}
\label{lab:methodology}
\begin{figure*}[!ht]
        \centering
        \includegraphics[width=0.8\linewidth]{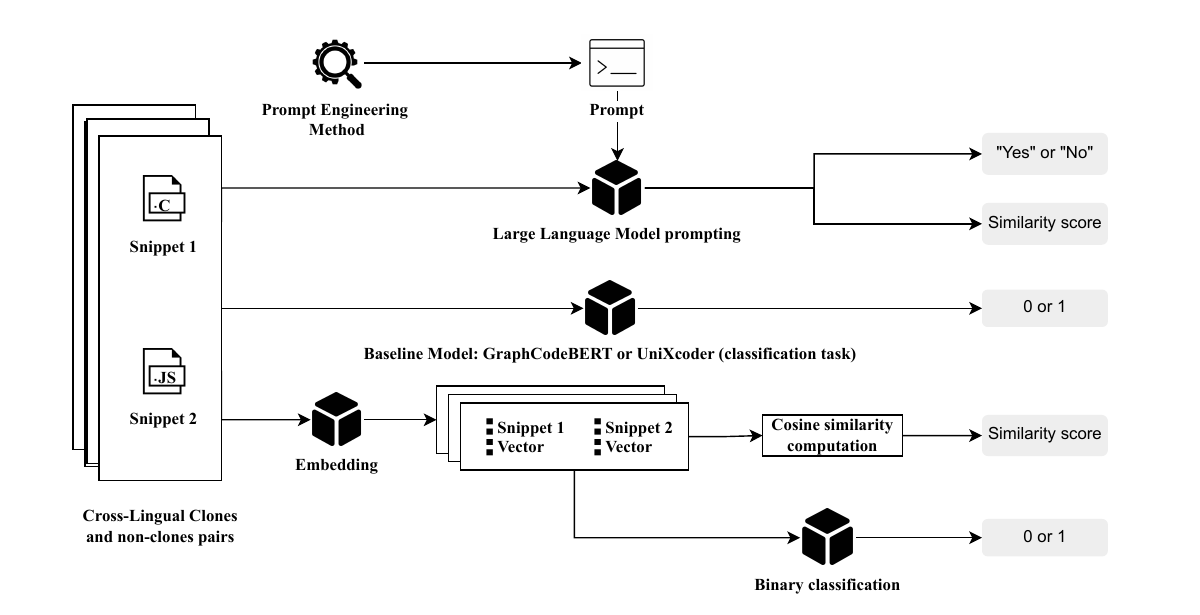} 
        \vspace{-0.5cm}
    \caption{Experimental workflow for cross-lingual code clone detection using LLMs vs Classification Models}
    \label{fig:workflow}
\end{figure*}

We overview the experimental methodology that is employed for assessing LLMs as well as baselines, and the classification models for the task of cross-lingual code clone detection. Figure~\ref{fig:workflow} illustrates the process, highlighting the input datasets,  the models as well as the expected outputs.

\subsubsection{Cross-Lingual Code Clone Detection as an NLP Task.} \hfill\\

We investigate the potential of LLMs for cross-lingual code clone detection using prompt engineering to explore the highest potential that LLMs can provide for this task. 

We build on zero-shot and chain-of-thought prompting techniques and design eight prompts which, each include a specially-crafted instruction as well as the raw content of the pair of code fragments to analyse for clone detection.  The expected output format varies depending on the prompt design: it can be either a straightforward binary "yes/no" answer or a numerical value representing a similarity score.

The proposed experimental methodology aims to assess the LLMs' ability for actual analysis of code semantics across different programming languages. Ultimately, we can provide measurements on the efficacy of LLMs for accurate cross-lingual code clone detection.

\noindent
{\bf LLM Prompting.} Our experiments are considering several prompts designed following two strategies: Zero-shot prompting, which is a classical and baseline approach used in the current literature~\cite{du_adaccd_2024} and Chain-of-Thought, which is a state-of-the-art prompting strategy that has been shown successful for several software engineering tasks~\cite{dou_towards_2023}. All prompts are provided in \Cref{tab:prompt}.

\begin{table*}[!ht]
    \centering
    \caption{List of Prompts Designed to Assess the Effectiveness of LLMs for the Task of Cross-Lingual Code Clone Detection}
    \label{tab:prompt}

\begin{subtable}[t]{1\linewidth}
\centering
\caption{Initial Pre-designed Prompts}
\resizebox{1\linewidth}{!}{%
\begin{threeparttable}
\begin{tabular}{>{\small}l>{\small}l>{\small}p{12cm}>{\small}c}

    \toprule
    \textbf{Prompting} &\textbf{Name} & \textbf{Prompt} & \textbf{output} \\ \midrule

    \textbf{Zero-shot} & Simple prompt & Analyze the following two code snippets and determine whether they are clones, regardless of the programming language. Respond with ‘yes’ if the code snippets are clones or ‘no’ if not. & yes/no \\ \midrule
    
    \multirow{6}{*}{\vspace{-5cm} \textbf{Chain of thought}} 
    &Similar line & Analyze the following two code snippets for code clone detection, regardless of the programming language. You should first report which lines of code are more similar. Then based on the report, please answer whether these two codes are a clone pair. The response should be ‘yes’ or ‘no’. & yes/no\\ \cline{2-4}

    & Reasoning& Provide a detailed reasoning process for detecting code clones in the following two code snippets, regardless of the programming language. Based on your analysis, respond with ‘yes’ if the code snippets are clones or ‘no’ if they are not. & yes/no\\ \cline{2-4}

   &Integrate & Analyze the following two code snippets to assess their similarity and determine if they are code clones, regardless of the programming language. Provide a similarity score between 0 and 10, where a higher score indicates more similarity. Additionally, presents a detailed reasoning process for detecting code clones. Conclude by ‘yes’ if they are clones or ‘no’ otherwise. & yes/no\\ \cline{2-4}
    
    & \multirow{2}{*}{Separate code} & \textbf{Step 1 :} Analyze the following code snippet and explain the function of the snippet.\\  
    & & \textbf{Step 2 :}  Analyze the following functions of two code snippets and determine if they are code clones, regardless of the programming language. The function of the first code snippet is \{step 1 output\} and the function of the second is \{step 1 output\}. Please answer ‘yes’ if the code snippets are clones, regardless of the programming language, or ‘no’ if they are not. &  yes/no \\ \cline{2-4}
    & \multirow{2}{*}{Separate Explanation} & \textbf{Step 1 :} Similarity/Reasoning/Difference/Integrated process without the cloning conclusion. \\ 
    & & \textbf{Step 2 :} Analyze the following two code snippets and determine if they are code clones. The Clone Similarity/Reasoning/Difference Integrated information of the first and the second code is: Please respond with ‘yes’ if the code snippets are clones or ‘no’ if they are not. &  yes/no \\ \cline{2-4}

    & Code similarity & Assess the similarity of the following two code snippets and provide a similarity score between 0 and 10. A higher score indicates that the two codes are more similar. Output the similarity score. & Similarity score\\ \bottomrule

\end{tabular}
\vspace{0.5cm}

\end{threeparttable}
}
\end{subtable}

\begin{subtable}[t]{1\linewidth}
\centering
\caption{Improved Prompt - Explicitly Describing the expected meaning of "Clone" to the LLM for cross-lingual clone detection}
\label{tab:prompt-simple-prompt-2}
\resizebox{1\linewidth}{!}{%
\begin{threeparttable}
\begin{tabular}{>{\small}l>{\small}l>{\small}p{12cm}>{\small}c}
    \toprule
    \textbf{Prompting} &\textbf{Name} & \textbf{Prompt} & \textbf{output} \\ \midrule

    \textbf{Zero-shot} &Improved simple prompt & Consider the overall structure and logic of the following two codes and determine if the two code snippets perform a similar task. Respond with ‘yes’ if the two codes perform similar tasks or ‘no’ otherwise. &  yes/no \\ \midrule

\end{tabular}

\end{threeparttable}}
\end{subtable}
    
\end{table*}

\vspace{0.5cm}

\subsubsection{Cross-Lingual Code Clone Detection as a Classification Task} \hfill\\

\textbf{Text-embedding-3-large Model.}  
We investigate, under the same conditions as LLMs, the potential for traditional classification mechanisms based on machine learning models. To that end, we follow the steps used by Keller et al.~\cite{keller_what_2022} in their learning-based approach for code clone detection. In their approach a classifier is trained based on sample sets of clone and non-clone pairs where each pair is represented by the absolute difference between the embedding vectors of each code fragment in the pair. Compared to other concatenation techniques that were assessed, their work conclude that differencing achieved the best results. In our case, however, the representations are yielded Embedding models that are provided alongside the LLM infrastructure of OpenAI. We use Text-embedding-3-large \footnote{https://openai.com/index/new-and-improved-embedding-model/} embedding model, which was trained on massive code and text data to yield representations that maximally capture semantics and can serve for similarity analysis. Finally, we train a binary classifier based on the yielded representations and using two basic learners: k Nearest Neighbors (k-NN) and Support Vector Machines (SVM).

Beyond the learning-based classification approach, we also investigate a straightforward approach where we compute the similarity between two fragments in a cross-lingual pair represented in a single representation space. To that end, we apply the cosine similarity metric on the Ada embeddings and check for the optimal threshold value to achieve the highest performance in terms of clone pair identification.

\textbf{ Baseline Models.}
We fine-tune the selected baseline models to optimize their performance for the task. These fine-tuned baselines are then systematically compared to the large language models (LLMs) to evaluate their relative effectiveness. This comparison allows us to assess how traditional models perform compared to the LLMs across the same task, providing insights into the strengths and limitations of each approach. 

\section{Evaluation} 
\label{sec:evaluation}

For each research question presented in \ref{rqs}, we first present the experiment we conducted and then show the result.

 \begin{table}[h]
    \centering
        \caption{Performance of GPT-3.5.-Turbo with Various Prompts on the Task of Cross-lingual Code Clone Detection}
          \resizebox{0.7\linewidth}{!}{
            \begin{tabular}{l|l|ccc|ccc}
            \toprule
                \textbf{} & \multicolumn{4}{c}{\textbf{XLCoST}}& \multicolumn{3}{c}{\textbf{CodeNet}}\\
                \midrule
                
                \textbf{Prompt name}        & \textbf{Clone type}           & \textbf{Recall}    & \textbf{Precision} &\textbf{ F1-score} &  \textbf{Recall }    & \textbf{Precision} & \textbf{F1-score}\\ \midrule
                
                \multirow{2}{*} \textbf{Simple prompt}
                & Non-clone         & {0.99}      & {0.97}      & {0.98}          & 1.00      & 0.56       & 0.72\\ 
                & Clone             & {0.97}      & {0.99}      & {0.98}          & 0.21      & 1.00       & 0.34\\ \midrule

                \multirow{2}{*} \textbf{Khajezade et al. ~\cite{khajezade_investigating_2024}}
                & Non-clone         & {-}      & {-}      & {-}          & 1.00      & 0.56       & 0.72\\ 
                & Clone             & {-}      & {-}      & {-}          & 0.21      & 1.00       & 0.35\\ \midrule
                
                \multirow{2}{*} \textbf{Similar line}
                & Non-clone         & {1.00}      & {0.98}      & \cellcolor{gray!20}{0.99}          & 1.00      & 0.60       & 0.75\\ 
                & Clone             & {0.98}      & {1.00}      & \cellcolor{gray!20}{0.99}          & 0.33      & {0.99}       & 0.49\\ \midrule
                
                \multirow{2}{*} \textbf{Reasoning }             
                & Non-clone         & {0.98}      & {0.99}      & {0.98}          & 0.99      & {0.71}       & {0.82}\\
                & Clone             & {0.99}      & {0.98}      & {0.98}          & {0.59}      & 0.98       &{0.74}\\ \midrule
                
                \multirow{2}{*} \textbf{Integrate}
                & Non-clone         & {0.99}      & {0.90}      & {0.94}          & 0.99      & 0.60       & 0.74\\ 
                & Clone             & {0.89}      & {0.99}      & {0.93}          & 0.32      & 0.98       & 0.49\\ \midrule
                
                \multirow{2}{*} \textbf{Separate code}           
                & Non-clone         & {1.00}      & {0.98}      & {0.99}          & 1.00      & {0.58}       & {0.73}\\ 
                & Clone             & {0.98}      & {1.00}      & {0.99}          & {0.27}      & {1.00}       & {0.42}\\ \midrule
                
                \multirow{2}{*} \textbf{Separate explanations}   
                & Non-clone         & {0.98}      & {0.99}      & {0.98}          & {0.99}      & 0.73      & \cellcolor{gray!20}{0.84}\\ 
                & Clone             & {0.99}      & {0.98}      & {0.98}          & 0.64      & 0.99       & \cellcolor{gray!20}{0.78}\\ 
                \bottomrule
                
            \end{tabular}
    \label{tab:gpt3-5-results}}
\end{table}

\begin{table}[h]
      \centering
      \caption{Performance Comparison of LLMs on the task of Cross-Lingual Code Clone Detection - based on the "Simple Prompt"} 
      \resizebox{0.7\linewidth}{!}{
      \begin{adjustbox}{width=\linewidth}
        \begin{tabular}{l|l|ccc|ccc}
        \hline
             & \multicolumn{4}{c }{\textbf{XLCoST}}& \multicolumn{3}{ c }{\textbf{CodeNet}}\\
             \midrule
             \textbf{LLM} & \textbf{Clone type}      & \textbf{Recall}    & \textbf{Precision}     & \textbf{F1-score}  & \textbf{Recall}     & \textbf{Precision}     & \textbf{F1-score}\\ \midrule

            \multirow{2}{*}\textbf{Falcon-Instruct-7B}  
            & Non-clone & {0.15}    & {0.51}          & 0.23     & 0.11        & 0.48           & 0.17 \\ 
            & Clone &{0.86}    & {0.50}          & 0.63    & 0.89        & 0.50           & 0.64 \\ \midrule
            
             \multirow{2}{*}\textbf{StarChat-Beta-16B}   
            & Non-clone& 0.62      & 0.47          & 0.53     & 0.76      & 0.49          & 0.60\\ 
            & Clone & 0.30      & 0.45          & 0.36     & 0.22      & 0.48          & 0.30\\ \midrule
            
             \multirow{2}{*}\textbf{LLaMA2-Chat-7B}      
            & Non-clone& 0.01       & 1.00          & {0.01}     & 0.04        & 0.68          &0.08 \\ 
            & Clone & 1.00      & 0.50         & {0.67}    & 0.98        & 0.51          &\cellcolor{gray!20}0.67 \\ \midrule

            \multirow{2}{*}\textbf{Starcoder2-15b-instruct}   
            & Non-clone& 1.00      & 0.5          & 0.67     & 1.00      & 0.5          & 0.67\\ 
            & Clone & 0.01      & 1.00          & 0.03     & 0.02      & 0.96          & 0.03\\ \midrule
            
             \multirow{2}{*} \textbf{GPT-3.5-Turbo}       
            & Non-clone & {0.99}      & {0.97}          & \cellcolor{gray!20}{0.98}   & {1.00}        & {0.56}           & \cellcolor{gray!20}{0.72}\\ 
            & Clone & {0.97}      & {0.99}          & \cellcolor{gray!20}{0.98}    & {0.21}        & {1.00}           & {0.34}\\ \midrule
        \end{tabular}
    \label{tab:llms-comparison-simple-prompt1}

      \end{adjustbox}}
    
\end{table}

\subsection{[RQ.1] LLM Performance on Cross-Lingual Code Clone Detection and Impact of Prompt Engineering}
\label{rq1}
\noindent 
\textbf{Goal.} We aim to fill a gap in the literature with a comprehensive analysis of the effectiveness of LLMs on the task of cross-lingual code clone detection. In particular, we investigate the impact of state-of-the-art prompt engineering strategy on the performance of the LLMs: we thus experiment with Zero-shot and Chain-of-Thought (CoT)-based prompts. Depending on the desired output formats, the designed prompts fall into two categories:  

\begin{itemize} [leftmargin=*]
    \item output is a \textit{score [0-10]}: We design a prompt that instructs the models to generate a similarity score between 0 and 10 when given a pair of code fragments: the lowest score (0) means that the model finds no similarity between the fragments, while the highest score (10) means that the model finds that the similarity is perfect. This prompt is designed using the CoT prompting strategy.
    \item output is \textit{Yes/No}: We design several prompts to guide the model towards answering with "Yes" or "No" when given a pair of code fragments. One of these prompts is based on a straightforward Zero-Shot strategy, while we provide five (5) prompts building on Chain-of-Thought. The design of the prompts is varied to take into account the requested depth of analysis (line by line or overall block), the specification of need to provide a reasoning of the process, the breaking of the steps for decision, etc. This methodology is inspired by prior work~\cite{dou_towards_2023} leveraging LLMs for clone detection, which has not assessed cross-lingual capabilities.
\end{itemize}

\vspace{0.2cm}
\noindent 
\textbf{Experiments.}
 We evaluated the {five} LLMs considered in this study (cf. \Cref{lab:models}), namely GPT-3.5-Turbo, Starchat-$\beta$-16B, Falcon-7B-Instruct, LLAMA2-Chat-7B, and, {Starcoder2-15b-instruct} on the constructed subsets \Cref{lab:datasets}) from XLCoST and CodeNet. 
 We use the ChatGPT API to prompt the GPT-3.5-Turbo LLM and parse the outputs to compute the precision, recall, and F1 scores finally.
For prompts leading to outputs with a similarity score, we analyzed the LLMs' outputs across a range of threshold values for code clone decisions.
{Our experiments included a comparison of our prompts with a prior proposal for cross-language code clone detection, evaluated on our dataset.}

\noindent 
\textbf{Results.} 
\noindent Our experiments confirmed that many LLMs actually fail to support the Chain-of-Thought prompting strategy, confirming a claim by Dou et al.~\cite{dou_towards_2023}. Consequently, to analyze the performance of LLMs across different prompts, we focus on GPT-3.5-Turbo, which produced results when instructed with all our designed prompts. \Cref{tab:gpt3-5-results} summarizes the results for this LLM across all the "yes/no" output prompts.
On the XLCoST dataset, the CoT-based  "Separate explanations" prompt enables the model to achieve the highest {F1 score (98\% for non-clones and clones)}.
On the CodeNet dataset, the same prompt achieves 84\% and 78\% respectively for non-clones and clones. We can observe that the performance on the CodeNet dataset is lower than that of the XLCoST dataset.
This discrepancy in performance is due to the difference in complexity between the two datasets.
Indeed, XLCoST is mainly composed of source code implementing simple problems such as ``\textit{Compute modulus division by a power}'' while CodeNet is built based on coding challenges of higher complexity such as ``\textit{Given a string, find the minimum number of swaps needed to rearrange the characters into a palindrome}''.
Furthermore, some LLMs exhibit a clear difference in their ability to detect code clones compared to non-clones.
For instance, on the CodeNet dataset, the GPT-3.5-turbo model shows a 35 percentage point variation in the f1 score (72\% on non-clone \textit{vs.} 37\% on clones).
On the same dataset, we also observe a high variation of 47 for the `Falcon-Instruct-7B' (17\% on non-clones \textit{vs.} 64\% on clones) model.
Manual analysis revealed a tendency for this model to classify nearly all input code snippet pairs as clones, resulting in an inflated number of false positives.

\find{{\bf  [RQ-1.1] \ding{42} } In general, for simple and recurrent programming tasks, LLMs show a high ability to detect cross-lingual clones (e.g., XLCoST samples). However, when applied to code samples of challenging programming tasks (e.g., CodeNet), LLMs performance drop by $~20\%$ on average in terms of F1 score. This result is insightful as it calls for a different perspective in benchmarking cross-lingual code clone detection approaches. Indeed, while XLCost has been built for such purpose, CodeNet appears to offer a more realistic setting, in terms of complexity for assessment.}

The "Simple prompt" exhibits a relatively consistent performance across both datasets, suggesting its potential for broader applicability. A detailed performance comparison of all five LLMs using this prompt is presented in \Cref{tab:llms-comparison-simple-prompt1}. Notably, the GPT-3.5-turbo model outperformed other LLMs in this analysis.
{Also, our experiments demonstrate that our proposed simple prompt achieved comparable results to the previously suggested one.}

\find{{\bf  [RQ-1.2] \ding{42}} Compared to Zero-Short prompting, which is mainly used in the literature, Chain of Thought prompting improves the performance of LLMs by 1-28 percentage points in terms of overall F1-score. It is further noteworthy that GPT-3.5-Turbo substantially outperforms the other studied LLMs in the task of cross-lingual code clone detection.}

\begin{figure*}[!ht]
        \centering
        
        \begin{subfigure}{1\textwidth}
                \includegraphics[width=\textwidth]{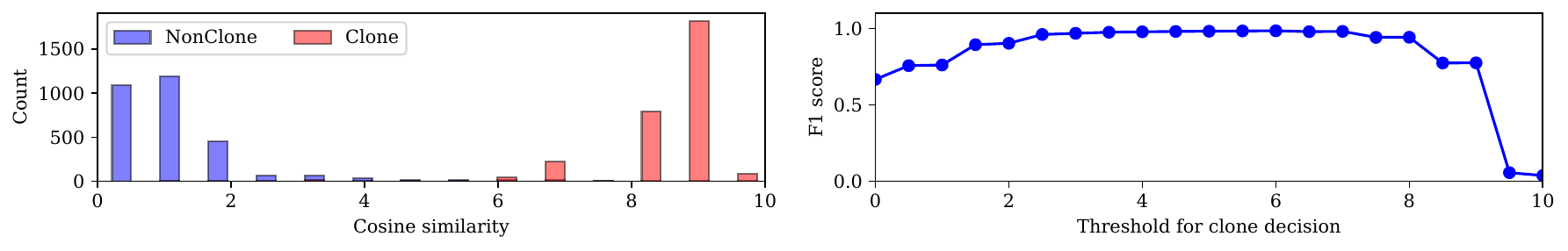}
                \caption{Results for the XLCoST dataset}
        \end{subfigure}
        
        \begin{subfigure}{1\textwidth}
                \centering
                \includegraphics[width=\textwidth]{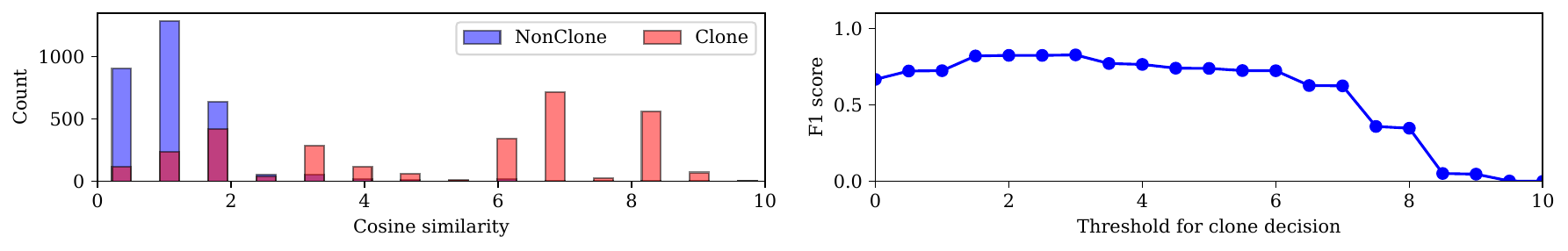} 
                \caption{Results for the CodeNet dataset}
        \end{subfigure}
        
        \caption{{Performance of GPT-3.5-Turbo prediction of Similarity Scores on the task of Cross-lingual Code Clone Detection}} 
        
        \label{fig:results-of-code-similarity-prompt}
\end{figure*}

In \Cref{fig:results-of-code-similarity-prompt}, we present the result of the GPT-3.5-turbo LLM with the "Code Similarity" prompt, designed to generate a similarity score between code snippets (0-10 scale)
The results show that with a threshold of 5, the LLM achieves an overall F1 score of  0.98 and 0.78 on the XLCoST and the CodeNet datasets, respectively.

\find{{\bf  [RQ-1.3] \ding{42}} Our experiments on the output requirements (Yes/No vs Similarity scores) underline that both approaches lead to similar performance scores. This suggests that the LLMs indeed consider the clone detection problem as a similarity computation task. }

\subsection{[RQ2] LLMs' Reasoning for Cross-Lingual Code Clone Detection}
\label{rq2}

\noindent
\textbf{Goal.} Our investigation of RQ.1 on the performance of LLMS for the cross-lingual code clone detection task has revealed, based on a manual analysis of some sample cases, that the code pairs are often obviously (for a human) a clone, and yet the LLM fails to identify it. In other cases, the pair is obviously (for a human) not a clone, and yet the LLM concludes that it is. We thus now aim to further study the extent to which the LLMs understand the task. Based on our analysis, notably of whether the LLMs understand what "clone" means in a cross-lingual setting, we propose an improved version of the Zero-shot prompt to validate that LLMs performance can be improved with a hint on the definition of clone.

\noindent
\textbf{Experiments.}
We qualitatively analyzed the outputs of all the experiments conducted in the first research question to collect insights in the failures of the LLMs for the task of cross-lingual code clone detection. 
Taking into account those results, we improve the Zero-shot prompt design and repeat the same experiments as in \Cref{rq1} based. In this RQ, we focus on the Zero-shot prompting, since it is a baseline, in order to better evaluate the impact of the improvement in the definition.

\noindent
\textbf{Results.}
Our qualitative review of LLMs' outputs reveals that Falcon-Instruct-7B and LLAMA2-Chat-7B consider, most of the time, any input pair of code fragments as clones, leading to an excessive amount of false positives.
Conversely, {Starcoder2-15b-instruct and} Starchat-$\beta$ considers most of the pairs as non-clones, yielding a significant number of false negatives.

The behavior of {Starcoder2-15b-instruct and} Starchat-$\beta$ are explained by the fact the reasoning of the LLM completely breaks down for the task of cross-lingual code clone detection. The LLM even goes far as to over-confidently state that {\em two code snippets of different programming languages cannot be clones}. Similar breakdowns in LLM reasoning have recently been documented in the literature~\cite{nezhurina2024alice}. Our experiments reveal for future research development in LLMs that cross-lingual code clone detection is a relevant and easy-to-assess task for checking potential reasoning breakdown of LLMs.
Unfortunately, {Starcoder2-15b-instruct and} Starchat-$\beta$ are not the only LLM with a breakdown in reasoning: Similar situations are also apparent in several outputs of GPT-3.5-turbo.

We analyze specifically the cases where the prompts require the outputs to include explanations from the Chain of Thought. A major finding was that in most of the cases where the LLM output was accurate, the LLMs actually stated that their decision was made taking into the \textbf{``overall structure and logic''} of the code fragments.
Based on this observation, we propose to design a new prompt, the `improved simple prompt,' that instructs the LLMs on what to consider as code clones. In other words, the improved prompt clearly defines the concept of code clones without using the term, which may be misleading to the LLMs: we instruct the model, in a Zero-Shot manner, to consider the overall structure and logic of the code snippets independently of their programming language (cf. \Cref{tab:prompt-simple-prompt-2}). 

\Cref{tab:llms-comparison-simple-prompt2} presents the performance achieved by the various LLMs based on the improved prompt. The performance improvement (compared to results when using a prompt that simply asks about "clones" - cf. \Cref{tab:llms-comparison-simple-prompt1}) is substantial. In particular, for  Starchat-$\beta$ and LLAMA2-Chat-7B, the new prompt helps the LLM to exhibit an improvement between {7 and 48} percentage points in terms of F1 score.  
 
\find{{\bf  [RQ-2] \ding{42}}
LLMs consistently are not fully aware of the notion of code clones in a cross-lingual setting. When prompted to adhere a definition of cloning as a measurement of similarity taking into account overall structure and logic, LLMs are able to demonstrate substantial success in the task of cross-lingual code clone detection.  On the XLCoST dataset, where cross-lingual clones are known (cf. RQ1) to be easier to detect, the improved prompt enables the LLaMA2-Chat-7B model to raise its F1 performance score from 0.01 to 0.80 for non-clones and 0.67 to 0.83 for clones, reducing the gap from GPT-3.5-Turbo by 14.5 percentage points.}
    \begin{table}[!t]
      \centering
      \caption{LLMs performance comparison with the improved prompt, designed based on LLMs common behavior}
      \resizebox{0.8\linewidth}{!}{
      \begin{adjustbox}{width=\linewidth}
        \begin{tabular}{l|l|ccc|ccc}
        \hline
             & \multicolumn{4}{c }{\textbf{XLCoST}}& \multicolumn{3}{ c }{\textbf{CodeNet}}\\
             \midrule
             \textbf{LLM} & \textbf{Clone type}      & \textbf{Recall}    & \textbf{Precision}     & \textbf{F1-score}  & \textbf{Recall}     & \textbf{Precision}     & \textbf{F1-score}\\ \midrule

            \multirow{2}{*}\textbf{Falcon-Instruct-7B}  
            & Non-clone                                  &{0.23}         & {0.54}          & 0.32                & 0.15          & 0.51           & 0.23 \\ 
            & Clone                                      & {0.80}        & 0.51            & 0.64                & 0.86           & 0.50           &{0.63}\\ \midrule
            
             \multirow{2}{*}\textbf{StarChat-Beta-16B}   
            & Non-clone                                  &{0.66}         & {0.60}          & 0.63                & 0.75          & 0.52           & 0.61 \\ 
            & Clone                                      & {0.57}        & 0.64            & 0.60                & 0.34           & 0.59           &{0.43}\\ \midrule

            \multirow{2}{*}\textbf{Starcoder2-15b-instruct}   
            & Non-clone                                  &{0.96}         & {0.58}          & 0.73                & 0.92          & 0.52           & 0.67 \\ 
            & Clone                                      & {0.31}        & 0.89            & 0.46                & 0.17           & 0.67           &{0.27}\\ \midrule
            
             \multirow{2}{*}\textbf{LLaMA2-Chat-7B}      
            & Non-clone                                  &{0.75}         & {0.87}          & 0.81                & 0.78          & 0.57           & 0.66 \\ 
            & Clone                                      & {0.89}        & 0.78            & 0.83                & 0.40           & 0.65           &{0.50}\\ \midrule
            
             \multirow{2}{*} \textbf{GPT-3.5-Turbo}       
            & Non-clone                                  &{0.98}         & {0.99}          & \cellcolor{gray!20}{0.98}                & 1.00          & 0.75           & \cellcolor{gray!20}{0.85} \\ 
            & Clone                                      & {0.99}        & 0.98            & \cellcolor{gray!20}{0.98}                & 0.66           & 1.00           &\cellcolor{gray!20}{0.79}\\ \midrule
        \end{tabular}
    \label{tab:llms-comparison-simple-prompt2}
      \end{adjustbox}}
    \end{table}

\subsection{[RQ3] Influence of the Programming Languages Syntactical Similarity on LLMs Performances }
\label{rq3}

\noindent
\textbf{Goal.} In the studied task, a clone pair involves two fragments from two distinct programming languages. Previous research questions focused on highlight the overall performance of LLMs on built pairs. We now attempt to check whether the specific combinations of programming languages may affect the performance of the LLM in identifying clones. In particular, we investigate whether the syntactic similarity of programming languages has an influence on the effectiveness of the LLMs. 

\noindent
\textbf{Experiments.}
Some programming languages share common core concepts, such as for controlling program flow (e.g., loops, conditionals, etc.) or storing data (e.g., variables). In terms of Syntax, we note for example that Java's syntax is most like C++ and C.  In fact, Java borrows heavily from C++ syntax, and therefore their programs may present similar structures. Similarly, languages like C\# (pronounced C-Sharp) share similarities with Java because they are object-oriented languages with similar features like classes, objects, inheritance, and encapsulation. To better highlight the potential combinations, we focus on the case of Java, and consider sets of pairs where the second programming language is varied across C, C\#, C++, Go, Ocalm, PHP, Python, Ruby, Rust, and JavaScript. 
We then disaggregate the performance metrics by programming language pair, focusing on experimental results with the CodeNet dataset from RQ-1 (cf. \Cref{rq1}). Given that we investigate the performance across the various prompt designs, we focus on the GPT-3.5-Turbo LLM, for which all prompt designs successfully returned answers.

\noindent 
\textbf{Results.}
\Cref{tab:results_by_programming_languages} shows all the results across various prompts and by programming language pair. 
We note that the performance difference is at least of $\sim$10 percentage points in terms of F1 score when the LLM is applied to fragments of Java-C language pair vs fragments of Java-Ocaml language pair. C is indeed very close to Java in terms of syntax and core concepts, while Ocaml employs different programming paradigms.

Nevertheless, the results also show that, while syntactic similarity of programming languages influences the performance of LLMs when instructed with a simple prompt, complex prompts with reasoning instructions (e.g., Reasoning) and the Improved simple prompt (which guides the LLM to consider logic) manages to reduce the performance gap for language pairs that are supposedly relatively different (e.g., Java-PHP).

\begin{table}[!t]
      \centering
      \caption{Performance of GPT-3.5-turbo  on the task of cross-lingual code clone detection - Detailed F1 scores by programming language pairs (Java - {\tt Lang-X} )} 
      \resizebox{0.8\linewidth}{!}{
        \begin{tabular}{>{\small}r|>{\small}c>{\small}c>{\small}c>{\small}c>{\small}c>{\small}c>{\small}c}
        \hline
        \textbf{\tt Lang-X}&\textbf{\thead{Simple\\Prompt}}& \textbf{\thead{Improved Simple\\Prompt}} & \textbf{\thead{Similar\\Line}} & \textbf{\thead{Reasoning}} & \textbf{\thead{Integrate}} & \textbf{\thead{Separate\\Explanation}} & \textbf{\thead{Separate\\Code}}  \\
        \hline
        \textbf{C}&              0.68&      0.85&      0.70&      0.83&      0.71&      0.84&      0.60 \\
        \textbf{C}\#&            0.54&      0.86&      0.65&      0.81&      0.66&      0.82&      0.60 \\
        \textbf{C++}&            0.59&      0.81&      0.66&      0.81&      0.67&      0.85&      0.56 \\
        \textbf{Go}&             0.46&      0.81&      0.57&      0.74&      0.61&      0.82&      0.54 \\
        \textbf{Ocaml}&          0.48&      0.81&      0.59&      0.73&      0.54&      0.75&      0.56 \\
        \textbf{PHP}&            0.50&      0.78&      0.62&      0.80&      0.64&      0.82&      0.59 \\
        \textbf{Python}&         0.52&      0.80&      0.59&      0.76&      0.61&      0.80&      0.59 \\
        \textbf{Ruby}&           0.49&      0.81&      0.59&      0.78&      0.52&      0.76&      0.57 \\
        \textbf{Rust}&           0.47&      0.86&      0.57&      0.75&      0.57&      0.80&      0.55 \\
        \textbf{JavaScript}&     0.53&      0.85&      0.63&      0.80&      0.63&      0.82&      0.56 \\
        \bottomrule
        \end{tabular}
    } 
    \label{tab:results_by_programming_languages}
    \end{table}

\find{{\bf  [RQ-3] \ding{42}} For the task of cross-lingual code clone detection, LLMs, when instructed with simple zero-shot prompts that do not clarify the definition of clones, see their performance influenced by the similarity of the languages of the code fragments in a pair. However, when the prompt explores the reasoning potential of the LLM, the differences in the programming language become less critical.
}

        \begin{figure*}[!htbp]
    \centering
    \begin{subfigure}{1\textwidth}
            \centering
            \includegraphics[scale=0.4]{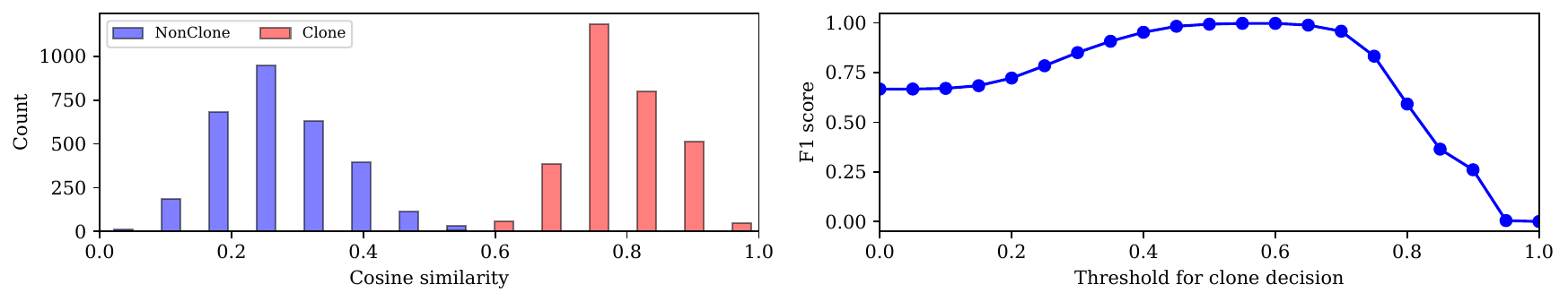}
            \caption{The performance of Text-embedding-3-large mixed with cosine similarity on XLCoST dataset}
    \end{subfigure}
    
    \begin{subfigure}{1\textwidth}
            \centering
            \includegraphics[scale=.4]{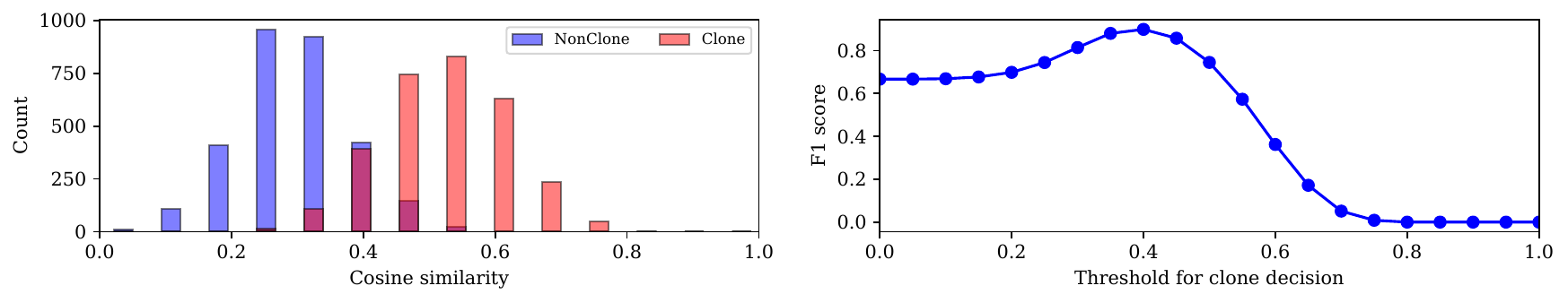} 
            \caption{The performance of Text-embedding-3-large mixed with cosine similarity on CodeNet dataset}
    \end{subfigure}
    
    \caption{In (a) and (b), we first measure the performance of Text-embedding-3-large mixed with cosine similarity at different thresholds for each dataset. Using a threshold of 0.5, we achieved an F1-score of 0.98 on the XLCOST dataset and 0.74 on the Codenet dataset.}
    \label{fig:ada_results_with_cosine}
    \end{figure*}

\begin{figure*}[!ht]
    \centering
    
    \begin{subfigure}{.9\textwidth}
            \centering
            \includegraphics[scale=0.6]{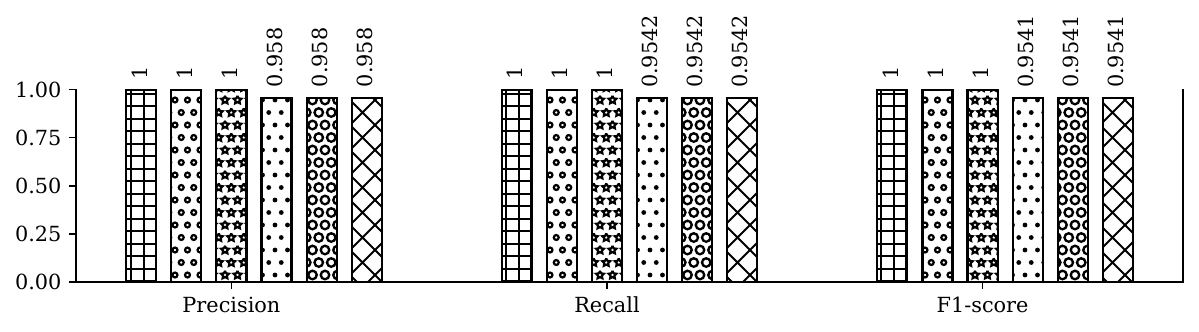}
            \vspace{-0.3cm}
            \caption{{Performance of the learning-based classifiers on the XLCoST dataset. 
            }}
    \end{subfigure}
    
    \begin{subfigure}{.9\textwidth}
            \centering
            \includegraphics[scale=0.6]{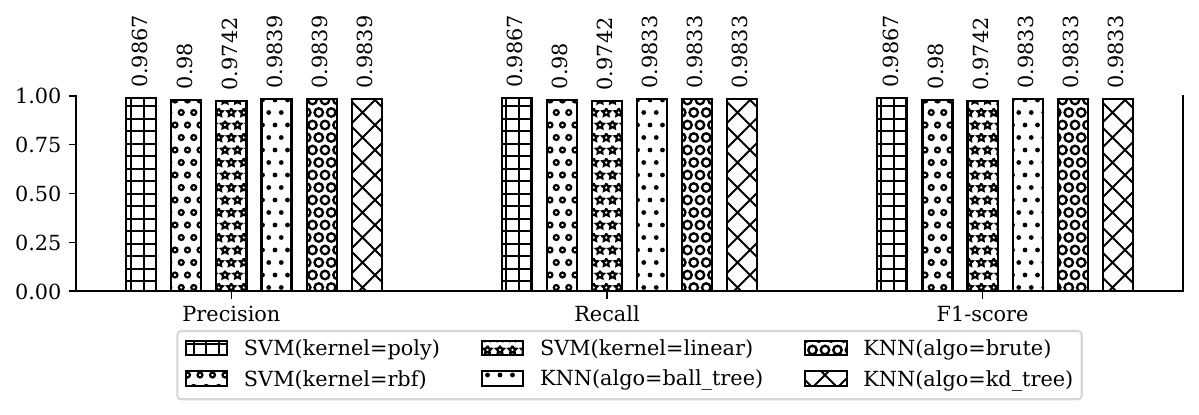} 
            \vspace{-0.3cm}
            \caption{Performance of the learning-based classifiers on the CodeNet dataset}
    \end{subfigure}
    
    \caption{{Impact of the binary classification on cross-lingual code clone detection using code embedding} }
        \label{fig:ada_global_results}
    
    \end{figure*}

\subsection{[RQ4] Binary Classifier \textit{vs.} Baselines \textit{vs.} LLMS}
\label{rq4}

\noindent
\textbf{Goal.} The primary objective of this research is to evaluate the performance of Embedding models in detecting cross-lingual code clones and to compare their effectiveness against large language models (LLMs). Code clone detection has been studied in the literature as a binary classification problem. Various techniques leveraging machine learning has then been adopted and showed promising results. We compare LLMs against implementations of such techniques. Since we aim to assess the added value of LLMs (beyond the representations), we build on the embedding model that is provided alongside the LLM by OpenAI. 

\noindent
\textbf{Experiments.}
We rely on the identified embedding model (Text-embedding-3-large - cf.~\Cref{lab:models}) to generate the representations of code fragments in our constructed benchmark. We conduct three separate experiments: 
\begin{enumerate}[leftmargin=*]
    \item  {\em straightforward similarity computation}: we use cosine similarity ~\cite{pedregosa2011scikit} to compute the distance between the embeddings of two code fragments in a given pair. We can identify an optimal threshold for code clone identification by considering varying threshold values. In this study, the threshold value is set to 5 for all evaluations.
    \item  {{\em Baselines model training for downstream classification task}: We conducted a series of experiments using two baseline models (GraphCodeBERT and UniXcoder). Given the input limitations of the baseline models, we made some adjustments to overcome the input length constraints of the baseline models when processing the dataset. To ensure robust evaluation, the selected dataset was split into an 80\% training set and a 20\% validation set. This experimental setup allowed us to fine-tune the baseline models and assess their ability to identify cross-lingual code clones under comparable conditions to the LLMs.}
    \item  {\em binary classifier training}: we consider some basic learners that we apply to the concatenation\footnote{cf. \Cref{lab:methodology} - following prior work~\cite{keller_what_2022}, we use the Absolute difference of embeddings as a concatenation method} of embeddings in code fragment pairs to train a binary classification model. As learners, we consider Support Vector Machines (SVM) ~\cite{cortes1995support} and k-Nearest Neighbors (k-NN)~\cite{cover1967nearest}. For SVM, we use three variants with  ball\_tree, kd\_tree, and brute algorithms, while for k-NN, we use four variants using the linear, polynomial, and rbf kernels. The experiments are performed using 10-Fold cross-validation on each dataset.
\end{enumerate} 

\noindent    
\textbf{Results.} \Cref{fig:ada_results_with_cosine} describes the performance (F1 score) that can be achieved when using cosine similarity thresholds for deciding on code clones. The threshold set at 5 permits to achieve up to  0.99, 0.97, and 0.98 scores regarding the overall recall, precision, and F1 for samples associated with the XLCoST dataset.  For the CodeNet dataset, the scores are  0.79, 0.84, and 0.78 for the overall recall, precision, and F1, respectively.

\Cref{tab:baselines-results} show the performance of the baseline models. They demonstrated superior performance compared to the LLMs, achieving an improvement of 15 percentage points in terms of the overall F1-score.
    \begin{table}[!h]
      \centering
      \caption{Comparison of LLM, Baselines, and Binary Classifier Performance}
      
      \resizebox{0.8\linewidth}{!}{
      \begin{adjustbox}{width=\linewidth}
        \begin{tabular}{l|l|l|ccc}
        \hline
              &  \multicolumn{4}{c}{\textbf{CodeNet}}\\
             \midrule
            Models & \textbf{Name}& \textbf{Clone type}      & \textbf{Recall}     & \textbf{Precision}     & \textbf{F1-score}\\ \midrule

            \multirow{6}{*}{Baselines}
           &  \multirow{3}{*}\textbf{GraphCodeBERT}  
             & Non-clone                                                  & 0.96          & 0.97           & 0.97 \\ 
            & & Clone                                                      & 0.98           & 0.97           &{0.97}\\ 
            &  &   Average                                                   & 0.97           & 0.97           &{0.97}\\ \cline{2-6}
            
            & \multirow{3}{*}\textbf{UnixCoder}   
             & Non-clone                                                 & 0.94          & 0.88           & 0.91 \\ 
            & & Clone                                                     & 0.89           & 0.95           &{0.92}\\ 
             &  &       Average                                              & 0.92           & 0.92           &{0.92}\\ \midrule

            LLM & \multirow{3}{*}\textbf{GPT-3.5-Turbo}   
             & Non-clone                                                 & 1.00          & 0.75           & 0.85 \\ 
            & & Clone                                                     & 0.66           & 1.00           &{0.79}\\ 
            &  &   Average                                                   & 0.83           & 0.88           &{0.82}\\ \midrule

            Binary classifier & \multirow{3}{*} \textbf{Text-embedding-3-large + SVM(kernel=`poly')}   
             & Non-clone                                                 & 0.99          & 0.98           & 0.99 \\ 
            & & Clone                                                     & 0.99           & 0.99           &{0.99}\\
            &  &   Average                                                   & 0.99           & 0.98           &\cellcolor{gray!20}{0.99}\\ \midrule
        \end{tabular}
    \label{tab:baselines-results}
      \end{adjustbox}}
    
    \end{table}

\Cref{fig:ada_global_results} summarizes the evaluation results of the trained classifiers. SVM achieves the highest scores at 1.00 for precision, recall, and F1. We further observe that k-NN achieves similar performance to SVM on the CodeNet dataset. This suggests homogeneity in the code samples enabling the learner to have a clear decision boundary.
In contrast, k-NN lags well behind SVM on the XLCoST dataset, suggesting higher diversity of samples (albeit potentially simple programs).

\find{{\bf  [RQ-4.1] \ding{42}}
The Text-embedding-3-large embedding model yields comprehensive representations of cross-lingual code fragments that enable rapidly identifying code clones based on straightforward similarity computation or learned classification.}

To provide a comprehensive understanding of our results, we present in \cref{tab:baselines-results} an in-depth comparison between traditional baselines, LLMs, and binary classifiers.

\find{{\bf  [RQ-4.2] \ding{42}}
 Compared to the performance achieved by baselines (GraphCodeBERT and UnixCoder), and LLMs instructed by prompts of varying complexity, we find that simple models (binary classifiers) achieve better results. This suggests that a major challenge in cross-lingual code clone detection is identifying a single representation space for all programming languages rather than attempting to have a universal reasoning engine.}

\section{Discussion} 
\label{sec:discussion}
This section explores the implications of our findings for cross-language code clone detection. Our investigation into the ability of LLMs to interpret cross-language code clones yielded valuable insights.

\noindent
\textbf{LLM Interpretation and Performance.}
Our experiments showed that LLMs can perform substantially better when we give them specific instructions about what to look for in the code (e.g., with the improved simple prompt) compared to when the prompt is generic and zero-shot based. The improved prompt, which instructs the LLM to reason about the overall structure and logic of the code leads LLMs to identify a large number of cross-lingual code clones (cf.~\Cref{tab:llms-comparison-simple-prompt2}). This finding suggests that LLMs can avoid dramatic reasoning failures when the prompt is sufficiently informative on the task: LLMs may need some guidance to see the bigger picture.

\noindent
\textbf{Representations vs. Prompt Engineering.}
The superiority of the learned classifier based on embedding models highlights the value of representations. Unlike prompts, which require careful design, an embedding model is a powerful, straightforward tool. It produces representations that capture the code's semantic meaning and functionality across different programming languages. Then based on the similarity of representations in a training set, a simple algorithm can learn decision boundaries, effectively overcoming language barriers in code clone detection. In contrast, prompts adds to the complexity and opacity of the reasoning workflow performed in an LLM.

\noindent
\textbf{Potential Data Leakage.}
Our experimental results using XLCoST demonstrate that the GPT-3.5-turbo exhibits a notable degree of familiarity with the samples. This may suggest potential leakage within this LLM training set.

\noindent
\textbf{Limitations and Threats to Validity.}
\ding{172} This study evaluated only {five} LLMs and {three} embedding models. Expanding the research to include a broader range of models with different architectures and training data could provide a more comprehensive understanding of LLM effectiveness in this task.
\ding{173} The data used in our evaluations are derived from widely used datasets in the literature. However, they may lack diversity, and this can lead to different observations on more diverse data (in terms of programming languages, complexity, ...).  
\ding{174} Further exploration of prompt design strategies is needed. This could involve investigating variations of Prompt 2, exploring entirely different prompt structures, and leveraging insights from this study to refine prompts for more optimal LLM performance.
\ding{175} Due to limitations in current LLM capabilities, most prompt approaches did not always provide definitive "yes/no" responses for clone detection. This necessitated manual verification and correction of results, which constitute an internal threat to validity. Future research can focus on developing prompt design or LLM training strategies that yield more reliable and automated outputs.
\ding{176} The Text-embedding-3-large model is likely based on an earlier architecture of the GPT family. Newer text embedding models may help achieve even better performance. 
\ding{177} The dataset used for fine-tuning GraphCodeBERT and UniXcoder was selected to align with both the input length constraints and the supported programming languages of these models. Consequently, the results reported in this work are limited to cases where input lengths do not exceed the maximum input size supported by the models, and the programming languages fall within the specific set supported by each model.
\ding{178} This study assumes a balanced dataset with an equal number of positive and negative pairs, which does not align with real-world conditions. In practical software projects, the vast majority of method pairs are not clones, resulting in a highly imbalanced scenario. While the balanced dataset used in this study provides insights into the techniques' effectiveness under controlled conditions, it may not fully represent their performance in real-world contexts. Future work should evaluate the studied techniques on datasets with more realistic imbalances to better understand their behavior and effectiveness in such scenarios.
\ding{179} The construction of negative pairs in the evaluation dataset introduces a potential threat to validity. While the authors employ an approach by distinguishing at the problem level to identify negative pairs, this design choice may inadvertently exclude certain types of negative pairs, particularly "close misses" pairs of code snippets that are similar but originate from distinct problems. As a result, the dataset may not fully capture the complexity of real-world cross-lingual clone detection scenarios, where such challenging distinctions are common.
Future work could address this by exploring alternative sampling strategies to include a broader variety of "close misses" negative pairs, and by examining how different definitions of "negative pairs" influence the evaluation outcomes. This would help create a dataset that more comprehensively reflects the diversity of challenges in real-world settings.

\section{Related Work}
\label{sec:related-work}
Prompt engineering is a challenging yet essential endeavor aimed at enhancing the performance of LLMs for specific tasks \cite{ye_prompt_2024}. Ekin et al.~\cite{ekin_prompt_2023} have demonstrated recently that to obtain accurate, relevant, and coherent output, users should design, refine, and optimize the prompt. Model understanding, domain specificity, and clarity are essential to identify the strengths and limits of the LLM, avoid ambiguity, and help it generate more accurate and relevant outputs. Our experimental results with a quick improvement of simple zero-shot prompts by providing a domain definition of the concept of clone confirms this finding for the field of software engineering.

Regarding code clone detection with LLMs, we can principally enumerate works by Dou et al.~\cite{dou_towards_2023} and Khajezade et al.~\cite{khajezade_investigating_2024}. In~\cite{dou_towards_2023}, the authors proposed an automated code clone detection for all types of clones with various prompt designs for a comprehensive empirical evaluation of several LLMs applied to clones from a single language (Java). In alignment with our findings, they show that LLMs of the GPT family outperform others and that introducing intermediate reasoning steps through a chain of thought improves the performance of LLMs for the clone detection task. 

Khajezade et al.~\cite{khajezade_investigating_2024} worked on mono and cross-lingual code clone detection using only GPT-3.5-Turbo (in contrast to our work, which explored various LLMs). They also used only zero-shot prompts and considered only Java-Ruby combinations (in contrast to our work, which considered 10 language pairs). Their study furthermore does not investigate the power of representations over prompts.

\section{Conclusion}
\label{sec:conclusion}
This study explored the potential of LLMs for cross-language code clone detection. We evaluated five LLMs and investigated how prompt engineering enhances their performance. Additionally, we examined the effectiveness of code representations in the identification of code clones. Overall, our findings suggest the potential of using powerful LLMs for cross-language code clone detection. Nevertheless, While LLMs show promise, our results suggest that simpler binary classifiers leveraging robust language-agnostic representations can achieve higher, or at least comparable, performance for cross-language code clone detection.
The insights gained from this study provide a foundation for future research aimed at developing more robust LLM-based techniques and improving code representation methods to advance software engineering practices. Furthermore, the methodologies and approaches presented here offer valuable reference points for continued exploration in this emerging domain. Future work could focus on addressing dataset imbalances by simulating "close misses" or exploring alternative sampling strategies to construct more representative negative pairs, thereby enhancing the reliability and applicability of cross-language code clone detection techniques.

\section*{Data Availability}
\label{sec:data_availability}
To promote transparency and facilitate reproducibility, we make our artifacts available to the community at: 
\begin{center}
\url{https://github.com/TruX-DTF/CLCCD}
\end{center}
The repository includes the constructed benchmark of cross-lingual code clones, the prompt details, the experiment scripts, and the results.

\begin{acks}
This work was supported 
by (1) the Luxembourg Ministry of Foreign and European Affairs through their Digital4Development (D4D) portfolio under project LuxWAyS and (2) the European Research Council (ERC) under the European Union’s Horizon 2020 research and innovation programme (Project NATURAL - grant agreement N$^o$ 949014).
\end{acks}

\bibliographystyle{ACM-Reference-Format}
\bibliography{references}


\begin{thebibliography}{51}


\ifx \showCODEN    \undefined \def \showCODEN     #1{\unskip}     \fi
\ifx \showISBNx    \undefined \def \showISBNx     #1{\unskip}     \fi
\ifx \showISBNxiii \undefined \def \showISBNxiii  #1{\unskip}     \fi
\ifx \showISSN     \undefined \def \showISSN      #1{\unskip}     \fi
\ifx \showLCCN     \undefined \def \showLCCN      #1{\unskip}     \fi
\ifx \shownote     \undefined \def \shownote      #1{#1}          \fi
\ifx \showarticletitle \undefined \def \showarticletitle #1{#1}   \fi
\ifx \showURL      \undefined \def \showURL       {\relax}        \fi
\providecommand\bibfield[2]{#2}
\providecommand\bibinfo[2]{#2}
\providecommand\natexlab[1]{#1}
\providecommand\showeprint[2][]{arXiv:#2}

\bibitem[Almazrouei et~al\mbox{.}(2023)]%
        {falcon40b}
\bibfield{author}{\bibinfo{person}{Ebtesam Almazrouei}, \bibinfo{person}{Hamza
  Alobeidli}, \bibinfo{person}{Abdulaziz Alshamsi}, \bibinfo{person}{Alessandro
  Cappelli}, \bibinfo{person}{Ruxandra Cojocaru}, \bibinfo{person}{Merouane
  Debbah}, \bibinfo{person}{Etienne Goffinet}, \bibinfo{person}{Daniel Heslow},
  \bibinfo{person}{Julien Launay}, \bibinfo{person}{Quentin Malartic},
  \bibinfo{person}{Badreddine Noune}, \bibinfo{person}{Baptiste Pannier}, {and}
  \bibinfo{person}{Guilherme Penedo}.} \bibinfo{year}{2023}\natexlab{}.
\newblock \showarticletitle{{Falcon-40B}: an open large language model with
  state-of-the-art performance}.
\newblock  (\bibinfo{year}{2023}).
\newblock


\bibitem[Bai et~al\mbox{.}({[n.\,d.]})]%
        {baiexploring}
\bibfield{author}{\bibinfo{person}{Weiheng Bai}, \bibinfo{person}{Qiushi Wu},
  \bibinfo{person}{Kefu Wu}, {and} \bibinfo{person}{Kangjie Lu}.}
  \bibinfo{year}{[n.\,d.]}\natexlab{}.
\newblock \showarticletitle{Exploring the Influence of Prompts in LLMs for
  Security-Related Tasks}.
\newblock  (\bibinfo{year}{[n.\,d.]}).
\newblock


\bibitem[Cheng et~al\mbox{.}(2017)]%
        {cheng_clcminer_2017}
\bibfield{author}{\bibinfo{person}{Xiao Cheng}, \bibinfo{person}{Zhiming Peng},
  \bibinfo{person}{Lingxiao Jiang}, \bibinfo{person}{Hao Zhong},
  \bibinfo{person}{Haibo Yu}, {and} \bibinfo{person}{Jianjun Zhao}.}
  \bibinfo{year}{2017}\natexlab{}.
\newblock \showarticletitle{\textit{{CLCMiner}}: {Detecting} {Cross}-{Language}
  {Clones} without {Intermediates}}.
\newblock \bibinfo{journal}{\emph{IEICE Trans. Inf. \& Syst.}}
  \bibinfo{volume}{E100.D}, \bibinfo{number}{2} (\bibinfo{year}{2017}),
  \bibinfo{pages}{273--284}.
\newblock
\showISSN{0916-8532, 1745-1361}
\href{https://doi.org/10.1587/transinf.2016EDP7334}{doi:\nolinkurl{10.1587/transinf.2016EDP7334}}


\bibitem[Cortes and Vapnik(1995)]%
        {cortes1995support}
\bibfield{author}{\bibinfo{person}{Corinna Cortes} {and}
  \bibinfo{person}{Vladimir Vapnik}.} \bibinfo{year}{1995}\natexlab{}.
\newblock \showarticletitle{Support-vector networks}.
\newblock \bibinfo{journal}{\emph{Machine learning}}  \bibinfo{volume}{20}
  (\bibinfo{year}{1995}), \bibinfo{pages}{273--297}.
\newblock


\bibitem[Cover and Hart(1967)]%
        {cover1967nearest}
\bibfield{author}{\bibinfo{person}{Thomas Cover} {and} \bibinfo{person}{Peter
  Hart}.} \bibinfo{year}{1967}\natexlab{}.
\newblock \showarticletitle{Nearest neighbor pattern classification}.
\newblock \bibinfo{journal}{\emph{IEEE transactions on information theory}}
  \bibinfo{volume}{13}, \bibinfo{number}{1} (\bibinfo{year}{1967}),
  \bibinfo{pages}{21--27}.
\newblock


\bibitem[Dou et~al\mbox{.}(2023)]%
        {dou_towards_2023}
\bibfield{author}{\bibinfo{person}{Shihan Dou}, \bibinfo{person}{Junjie Shan},
  \bibinfo{person}{Haoxiang Jia}, \bibinfo{person}{Wenhao Deng},
  \bibinfo{person}{Zhiheng Xi}, \bibinfo{person}{Wei He},
  \bibinfo{person}{Yueming Wu}, \bibinfo{person}{Tao Gui},
  \bibinfo{person}{Yang Liu}, {and} \bibinfo{person}{Xuanjing Huang}.}
  \bibinfo{year}{2023}\natexlab{}.
\newblock \bibinfo{title}{Towards {Understanding} the {Capability} of {Large}
  {Language} {Models} on {Code} {Clone} {Detection}: {A} {Survey}}.
\newblock
\urldef\tempurl%
\url{http://arxiv.org/abs/2308.01191}
\showURL{%
\tempurl}


\bibitem[Du et~al\mbox{.}(2024)]%
        {du_adaccd_2024}
\bibfield{author}{\bibinfo{person}{Yangkai Du}, \bibinfo{person}{Tengfei Ma},
  \bibinfo{person}{Lingfei Wu}, \bibinfo{person}{Xuhong Zhang}, {and}
  \bibinfo{person}{Shouling Ji}.} \bibinfo{year}{2024}\natexlab{}.
\newblock \bibinfo{title}{{AdaCCD}: {Adaptive} {Semantic} {Contrasts}
  {Discovery} {Based} {Cross} {Lingual} {Adaptation} for {Code} {Clone}
  {Detection}}.
\newblock
\urldef\tempurl%
\url{http://arxiv.org/abs/2311.07277}
\showURL{%
\tempurl}


\bibitem[Ekin(2023)]%
        {ekin_prompt_2023}
\bibfield{author}{\bibinfo{person}{Sabit Ekin}.}
  \bibinfo{year}{2023}\natexlab{}.
\newblock \bibinfo{title}{Prompt {Engineering} {For} {ChatGPT}: {A} {Quick}
  {Guide} {To} {Techniques}, {Tips}, {And} {Best} {Practices}}.
\newblock
\href{https://doi.org/10.36227/techrxiv.22683919.v2}{doi:\nolinkurl{10.36227/techrxiv.22683919.v2}}


\bibitem[El~Arnaoty and Servant(2024)]%
        {el2024onespace}
\bibfield{author}{\bibinfo{person}{Mohammed El~Arnaoty} {and}
  \bibinfo{person}{Francisco Servant}.} \bibinfo{year}{2024}\natexlab{}.
\newblock \showarticletitle{OneSpace: Detecting cross-language clones by
  learning a common embedding space}.
\newblock \bibinfo{journal}{\emph{Journal of Systems and Software}}
  \bibinfo{volume}{208} (\bibinfo{year}{2024}), \bibinfo{pages}{111911}.
\newblock


\bibitem[Fang et~al\mbox{.}(2023)]%
        {fang_tcccd_2023}
\bibfield{author}{\bibinfo{person}{Yong Fang}, \bibinfo{person}{Fangzheng
  Zhou}, \bibinfo{person}{Yijia Xu}, {and} \bibinfo{person}{Zhonglin Liu}.}
  \bibinfo{year}{2023}\natexlab{}.
\newblock \showarticletitle{{TCCCD}: {Triplet}-{Based} {Cross}-{Language}
  {Code} {Clone} {Detection}}.
\newblock \bibinfo{journal}{\emph{Applied Sciences}} \bibinfo{volume}{13},
  \bibinfo{number}{21} (\bibinfo{date}{Nov.} \bibinfo{year}{2023}),
  \bibinfo{pages}{12084}.
\newblock
\showISSN{2076-3417}
\href{https://doi.org/10.3390/app132112084}{doi:\nolinkurl{10.3390/app132112084}}


\bibitem[Gu et~al\mbox{.}(2024)]%
        {gu2024cruxeval}
\bibfield{author}{\bibinfo{person}{Alex Gu}, \bibinfo{person}{Baptiste
  Rozi{\`e}re}, \bibinfo{person}{Hugh Leather}, \bibinfo{person}{Armando
  Solar-Lezama}, \bibinfo{person}{Gabriel Synnaeve}, {and}
  \bibinfo{person}{Sida~I Wang}.} \bibinfo{year}{2024}\natexlab{}.
\newblock \showarticletitle{Cruxeval: A benchmark for code reasoning,
  understanding and execution}.
\newblock \bibinfo{journal}{\emph{arXiv preprint arXiv:2401.03065}}
  (\bibinfo{year}{2024}).
\newblock


\bibitem[Guo et~al\mbox{.}(2022a)]%
        {guo_unixcoder_2022}
\bibfield{author}{\bibinfo{person}{Daya Guo}, \bibinfo{person}{Shuai Lu},
  \bibinfo{person}{Nan Duan}, \bibinfo{person}{Yanlin Wang},
  \bibinfo{person}{Ming Zhou}, {and} \bibinfo{person}{Jian Yin}.}
  \bibinfo{year}{2022}\natexlab{a}.
\newblock \bibinfo{title}{{UniXcoder}: {Unified} {Cross}-{Modal} {Pre}-training
  for {Code} {Representation}}.
\newblock
\urldef\tempurl%
\url{http://arxiv.org/abs/2203.03850}
\showURL{%
\tempurl}


\bibitem[Guo et~al\mbox{.}(2022b)]%
        {guo2022unixcoder}
\bibfield{author}{\bibinfo{person}{Daya Guo}, \bibinfo{person}{Shuai Lu},
  \bibinfo{person}{Nan Duan}, \bibinfo{person}{Yanlin Wang},
  \bibinfo{person}{Ming Zhou}, {and} \bibinfo{person}{Jian Yin}.}
  \bibinfo{year}{2022}\natexlab{b}.
\newblock \showarticletitle{Unixcoder: Unified cross-modal pre-training for
  code representation}.
\newblock \bibinfo{journal}{\emph{arXiv preprint arXiv:2203.03850}}
  (\bibinfo{year}{2022}).
\newblock


\bibitem[Guo et~al\mbox{.}(2020)]%
        {guo2020graphcodebert}
\bibfield{author}{\bibinfo{person}{Daya Guo}, \bibinfo{person}{Shuo Ren},
  \bibinfo{person}{Shuai Lu}, \bibinfo{person}{Zhangyin Feng},
  \bibinfo{person}{Duyu Tang}, \bibinfo{person}{Shujie Liu},
  \bibinfo{person}{Long Zhou}, \bibinfo{person}{Nan Duan},
  \bibinfo{person}{Alexey Svyatkovskiy}, \bibinfo{person}{Shengyu Fu},
  {et~al\mbox{.}}} \bibinfo{year}{2020}\natexlab{}.
\newblock \showarticletitle{Graphcodebert: Pre-training code representations
  with data flow}.
\newblock \bibinfo{journal}{\emph{arXiv preprint arXiv:2009.08366}}
  (\bibinfo{year}{2020}).
\newblock


\bibitem[Guo et~al\mbox{.}(2021)]%
        {guo_graphcodebert_2021}
\bibfield{author}{\bibinfo{person}{Daya Guo}, \bibinfo{person}{Shuo Ren},
  \bibinfo{person}{Shuai Lu}, \bibinfo{person}{Zhangyin Feng},
  \bibinfo{person}{Duyu Tang}, \bibinfo{person}{Shujie Liu},
  \bibinfo{person}{Long Zhou}, \bibinfo{person}{Nan Duan},
  \bibinfo{person}{Alexey Svyatkovskiy}, \bibinfo{person}{Shengyu Fu},
  \bibinfo{person}{Michele Tufano}, \bibinfo{person}{Shao~Kun Deng},
  \bibinfo{person}{Colin Clement}, \bibinfo{person}{Dawn Drain},
  \bibinfo{person}{Neel Sundaresan}, \bibinfo{person}{Jian Yin},
  \bibinfo{person}{Daxin Jiang}, {and} \bibinfo{person}{Ming Zhou}.}
  \bibinfo{year}{2021}\natexlab{}.
\newblock \bibinfo{title}{{GraphCodeBERT}: {Pre}-training {Code}
  {Representations} with {Data} {Flow}}.
\newblock
\urldef\tempurl%
\url{http://arxiv.org/abs/2009.08366}
\showURL{%
\tempurl}


\bibitem[Jiang et~al\mbox{.}(2007)]%
        {jiang_deckard_2007}
\bibfield{author}{\bibinfo{person}{Lingxiao Jiang}, \bibinfo{person}{Ghassan
  Misherghi}, \bibinfo{person}{Zhendong Su}, {and} \bibinfo{person}{Stephane
  Glondu}.} \bibinfo{year}{2007}\natexlab{}.
\newblock \showarticletitle{{DECKARD}: {Scalable} and {Accurate} {Tree}-{Based}
  {Detection} of {Code} {Clones}}. In \bibinfo{booktitle}{\emph{29th
  {International} {Conference} on {Software} {Engineering} ({ICSE}'07)}}.
  \bibinfo{publisher}{IEEE}, \bibinfo{address}{Minneapolis, MN, USA},
  \bibinfo{pages}{96--105}.
\newblock
\showISBNx{978-0-7695-2828-1}
\href{https://doi.org/10.1109/ICSE.2007.30}{doi:\nolinkurl{10.1109/ICSE.2007.30}}


\bibitem[Juergens et~al\mbox{.}(2009)]%
        {juergens_code_2009}
\bibfield{author}{\bibinfo{person}{Elmar Juergens}, \bibinfo{person}{Florian
  Deissenboeck}, \bibinfo{person}{Benjamin Hummel}, {and}
  \bibinfo{person}{Stefan Wagner}.} \bibinfo{year}{2009}\natexlab{}.
\newblock \showarticletitle{Do code clones matter?}. In
  \bibinfo{booktitle}{\emph{2009 {IEEE} 31st {International} {Conference} on
  {Software} {Engineering}}}. \bibinfo{publisher}{IEEE},
  \bibinfo{address}{Vancouver, BC, Canada}, \bibinfo{pages}{485--495}.
\newblock
\showISBNx{978-1-4244-3453-4}
\href{https://doi.org/10.1109/ICSE.2009.5070547}{doi:\nolinkurl{10.1109/ICSE.2009.5070547}}


\bibitem[Kamiya et~al\mbox{.}(2002)]%
        {kamiya_ccfinder_2002}
\bibfield{author}{\bibinfo{person}{T. Kamiya}, \bibinfo{person}{S. Kusumoto},
  {and} \bibinfo{person}{K. Inoue}.} \bibinfo{year}{2002}\natexlab{}.
\newblock \showarticletitle{{CCFinder}: a multilinguistic token-based code
  clone detection system for large scale source code}.
\newblock \bibinfo{journal}{\emph{IIEEE Trans. Software Eng.}}
  \bibinfo{volume}{28}, \bibinfo{number}{7} (\bibinfo{date}{July}
  \bibinfo{year}{2002}), \bibinfo{pages}{654--670}.
\newblock
\showISSN{0098-5589}
\href{https://doi.org/10.1109/TSE.2002.1019480}{doi:\nolinkurl{10.1109/TSE.2002.1019480}}


\bibitem[Keller et~al\mbox{.}(2022)]%
        {keller_what_2022}
\bibfield{author}{\bibinfo{person}{Patrick Keller},
  \bibinfo{person}{Abdoul~Kader Kaboré}, \bibinfo{person}{Laura Plein},
  \bibinfo{person}{Jacques Klein}, \bibinfo{person}{Yves Le~Traon}, {and}
  \bibinfo{person}{Tegawendé~F. Bissyandé}.} \bibinfo{year}{2022}\natexlab{}.
\newblock \showarticletitle{What {You} {See} is {What} it {Means}! {Semantic}
  {Representation} {Learning} of {Code} based on {Visualization} and {Transfer}
  {Learning}}.
\newblock \bibinfo{journal}{\emph{ACM Trans. Softw. Eng. Methodol.}}
  \bibinfo{volume}{31}, \bibinfo{number}{2} (\bibinfo{date}{April}
  \bibinfo{year}{2022}), \bibinfo{pages}{1--34}.
\newblock
\showISSN{1049-331X, 1557-7392}
\href{https://doi.org/10.1145/3485135}{doi:\nolinkurl{10.1145/3485135}}


\bibitem[Khajezade et~al\mbox{.}(2024)]%
        {khajezade_investigating_2024}
\bibfield{author}{\bibinfo{person}{Mohamad Khajezade}, \bibinfo{person}{Jie~JW
  Wu}, \bibinfo{person}{Fatemeh~Hendijani Fard}, \bibinfo{person}{Gema
  Rodríguez-Pérez}, {and} \bibinfo{person}{Mohamed~Sami Shehata}.}
  \bibinfo{year}{2024}\natexlab{}.
\newblock \bibinfo{title}{Investigating the {Efficacy} of {Large} {Language}
  {Models} for {Code} {Clone} {Detection}}.
\newblock
\urldef\tempurl%
\url{http://arxiv.org/abs/2401.13802}
\showURL{%
\tempurl}


\bibitem[Kochhar et~al\mbox{.}(2016)]%
        {kochhar_large_2016}
\bibfield{author}{\bibinfo{person}{Pavneet~Singh Kochhar},
  \bibinfo{person}{Dinusha Wijedasa}, {and} \bibinfo{person}{David Lo}.}
  \bibinfo{year}{2016}\natexlab{}.
\newblock \showarticletitle{A {Large} {Scale} {Study} of {Multiple}
  {Programming} {Languages} and {Code} {Quality}}. In
  \bibinfo{booktitle}{\emph{2016 {IEEE} 23rd {International} {Conference} on
  {Software} {Analysis}, {Evolution}, and {Reengineering} ({SANER})}}.
  \bibinfo{publisher}{IEEE}, \bibinfo{address}{Suita},
  \bibinfo{pages}{563--573}.
\newblock
\showISBNx{978-1-5090-1855-0}
\href{https://doi.org/10.1109/SANER.2016.112}{doi:\nolinkurl{10.1109/SANER.2016.112}}


\bibitem[Koschke({[n.\,d.]})]%
        {koschke_survey_nodate}
\bibfield{author}{\bibinfo{person}{Rainer Koschke}.}
  \bibinfo{year}{[n.\,d.]}\natexlab{}.
\newblock \showarticletitle{Survey of {Research} on {Software} {Clones}}.
\newblock  (\bibinfo{year}{[n.\,d.]}).
\newblock


\bibitem[Li et~al\mbox{.}(2023b)]%
        {li2023zc}
\bibfield{author}{\bibinfo{person}{Jia Li}, \bibinfo{person}{Chongyang Tao},
  \bibinfo{person}{Zhi Jin}, \bibinfo{person}{Fang Liu}, {and}
  \bibinfo{person}{Ge Li}.} \bibinfo{year}{2023}\natexlab{b}.
\newblock \showarticletitle{ZC 3: Zero-Shot Cross-Language Code Clone
  Detection}. In \bibinfo{booktitle}{\emph{2023 38th IEEE/ACM International
  Conference on Automated Software Engineering (ASE)}}. IEEE,
  \bibinfo{pages}{875--887}.
\newblock


\bibitem[Li et~al\mbox{.}(2023a)]%
        {li_starcoder_2023}
\bibfield{author}{\bibinfo{person}{Raymond Li}, \bibinfo{person}{Loubna~Ben
  Allal}, \bibinfo{person}{Yangtian Zi}, \bibinfo{person}{Niklas Muennighoff},
  \bibinfo{person}{Denis Kocetkov}, \bibinfo{person}{Chenghao Mou},
  \bibinfo{person}{Marc Marone}, \bibinfo{person}{Christopher Akiki},
  \bibinfo{person}{Jia Li}, \bibinfo{person}{Jenny Chim}, \bibinfo{person}{Qian
  Liu}, \bibinfo{person}{Evgenii Zheltonozhskii}, \bibinfo{person}{Terry~Yue
  Zhuo}, \bibinfo{person}{Thomas Wang}, \bibinfo{person}{Olivier Dehaene},
  \bibinfo{person}{Mishig Davaadorj}, \bibinfo{person}{Joel Lamy-Poirier},
  \bibinfo{person}{João Monteiro}, \bibinfo{person}{Oleh Shliazhko},
  \bibinfo{person}{Nicolas Gontier}, \bibinfo{person}{Nicholas Meade},
  \bibinfo{person}{Armel Zebaze}, \bibinfo{person}{Ming-Ho Yee},
  \bibinfo{person}{Logesh~Kumar Umapathi}, \bibinfo{person}{Jian Zhu},
  \bibinfo{person}{Benjamin Lipkin}, \bibinfo{person}{Muhtasham Oblokulov},
  \bibinfo{person}{Zhiruo Wang}, \bibinfo{person}{Rudra Murthy},
  \bibinfo{person}{Jason Stillerman}, \bibinfo{person}{Siva~Sankalp Patel},
  \bibinfo{person}{Dmitry Abulkhanov}, \bibinfo{person}{Marco Zocca},
  \bibinfo{person}{Manan Dey}, \bibinfo{person}{Zhihan Zhang},
  \bibinfo{person}{Nour Fahmy}, \bibinfo{person}{Urvashi Bhattacharyya},
  \bibinfo{person}{Wenhao Yu}, \bibinfo{person}{Swayam Singh},
  \bibinfo{person}{Sasha Luccioni}, \bibinfo{person}{Paulo Villegas},
  \bibinfo{person}{Maxim Kunakov}, \bibinfo{person}{Fedor Zhdanov},
  \bibinfo{person}{Manuel Romero}, \bibinfo{person}{Tony Lee},
  \bibinfo{person}{Nadav Timor}, \bibinfo{person}{Jennifer Ding},
  \bibinfo{person}{Claire Schlesinger}, \bibinfo{person}{Hailey Schoelkopf},
  \bibinfo{person}{Jan Ebert}, \bibinfo{person}{Tri Dao},
  \bibinfo{person}{Mayank Mishra}, \bibinfo{person}{Alex Gu},
  \bibinfo{person}{Jennifer Robinson}, \bibinfo{person}{Carolyn~Jane Anderson},
  \bibinfo{person}{Brendan Dolan-Gavitt}, \bibinfo{person}{Danish Contractor},
  \bibinfo{person}{Siva Reddy}, \bibinfo{person}{Daniel Fried},
  \bibinfo{person}{Dzmitry Bahdanau}, \bibinfo{person}{Yacine Jernite},
  \bibinfo{person}{Carlos~Muñoz Ferrandis}, \bibinfo{person}{Sean Hughes},
  \bibinfo{person}{Thomas Wolf}, \bibinfo{person}{Arjun Guha},
  \bibinfo{person}{Leandro von Werra}, {and} \bibinfo{person}{Harm de Vries}.}
  \bibinfo{year}{2023}\natexlab{a}.
\newblock \bibinfo{title}{{StarCoder}: may the source be with you!}
\newblock
\urldef\tempurl%
\url{http://arxiv.org/abs/2305.06161}
\showURL{%
\tempurl}


\bibitem[Lin(2024)]%
        {lin2024write}
\bibfield{author}{\bibinfo{person}{Zhicheng Lin}.}
  \bibinfo{year}{2024}\natexlab{}.
\newblock \showarticletitle{How to write effective prompts for large language
  models}.
\newblock \bibinfo{journal}{\emph{Nature Human Behaviour}}
  (\bibinfo{year}{2024}), \bibinfo{pages}{1--5}.
\newblock


\bibitem[Lozhkov et~al\mbox{.}(2024)]%
        {starcoder2}
\bibfield{author}{\bibinfo{person}{Anton Lozhkov}, \bibinfo{person}{Raymond
  Li}, \bibinfo{person}{Loubna~Ben Allal}, \bibinfo{person}{Federico Cassano},
  \bibinfo{person}{Joel Lamy-Poirier}, \bibinfo{person}{Nouamane Tazi},
  \bibinfo{person}{Ao Tang}, \bibinfo{person}{Dmytro Pykhtar},
  \bibinfo{person}{Jiawei Liu}, \bibinfo{person}{Yuxiang Wei}, {et~al\mbox{.}}}
  \bibinfo{year}{2024}\natexlab{}.
\newblock \showarticletitle{StarCoder 2 and The Stack v2: The Next Generation}.
\newblock \bibinfo{journal}{\emph{arXiv preprint arXiv:2402.19173}}
  (\bibinfo{year}{2024}).
\newblock


\bibitem[Muennighoff et~al\mbox{.}(2022)]%
        {muennighoff2022crosslingual}
\bibfield{author}{\bibinfo{person}{Niklas Muennighoff}, \bibinfo{person}{Thomas
  Wang}, \bibinfo{person}{Lintang Sutawika}, \bibinfo{person}{Adam Roberts},
  \bibinfo{person}{Stella Biderman}, \bibinfo{person}{Teven~Le Scao},
  \bibinfo{person}{M~Saiful Bari}, \bibinfo{person}{Sheng Shen},
  \bibinfo{person}{Zheng-Xin Yong}, \bibinfo{person}{Hailey Schoelkopf},
  {et~al\mbox{.}}} \bibinfo{year}{2022}\natexlab{}.
\newblock \showarticletitle{Crosslingual generalization through multitask
  finetuning}.
\newblock \bibinfo{journal}{\emph{arXiv preprint arXiv:2211.01786}}
  (\bibinfo{year}{2022}).
\newblock


\bibitem[Nafi et~al\mbox{.}(2019)]%
        {nafi_clcdsa_2019}
\bibfield{author}{\bibinfo{person}{Kawser~Wazed Nafi},
  \bibinfo{person}{Tonny~Shekha Kar}, \bibinfo{person}{Banani Roy},
  \bibinfo{person}{Chanchal~K. Roy}, {and} \bibinfo{person}{Kevin~A.
  Schneider}.} \bibinfo{year}{2019}\natexlab{}.
\newblock \showarticletitle{{CLCDSA}: {Cross} {Language} {Code} {Clone}
  {Detection} using {Syntactical} {Features} and {API} {Documentation}}. In
  \bibinfo{booktitle}{\emph{2019 34th {IEEE}/{ACM} {International} {Conference}
  on {Automated} {Software} {Engineering} ({ASE})}}. \bibinfo{publisher}{IEEE},
  \bibinfo{address}{San Diego, CA, USA}, \bibinfo{pages}{1026--1037}.
\newblock
\showISBNx{978-1-72812-508-4}
\href{https://doi.org/10.1109/ASE.2019.00099}{doi:\nolinkurl{10.1109/ASE.2019.00099}}


\bibitem[Nakagawa et~al\mbox{.}(2021)]%
        {nakagawa_nil_2021}
\bibfield{author}{\bibinfo{person}{Tasuku Nakagawa}, \bibinfo{person}{Yoshiki
  Higo}, {and} \bibinfo{person}{Shinji Kusumoto}.}
  \bibinfo{year}{2021}\natexlab{}.
\newblock \showarticletitle{{NIL}: large-scale detection of large-variance
  clones}. In \bibinfo{booktitle}{\emph{Proceedings of the 29th {ACM} {Joint}
  {Meeting} on {European} {Software} {Engineering} {Conference} and {Symposium}
  on the {Foundations} of {Software} {Engineering}}}. \bibinfo{publisher}{ACM},
  \bibinfo{address}{Athens Greece}, \bibinfo{pages}{830--841}.
\newblock
\showISBNx{978-1-4503-8562-6}
\href{https://doi.org/10.1145/3468264.3468564}{doi:\nolinkurl{10.1145/3468264.3468564}}


\bibitem[Nezhurina et~al\mbox{.}(2024)]%
        {nezhurina2024alice}
\bibfield{author}{\bibinfo{person}{Marianna Nezhurina}, \bibinfo{person}{Lucia
  Cipolina-Kun}, \bibinfo{person}{Mehdi Cherti}, {and} \bibinfo{person}{Jenia
  Jitsev}.} \bibinfo{year}{2024}\natexlab{}.
\newblock \showarticletitle{Alice in Wonderland: Simple Tasks Showing Complete
  Reasoning Breakdown in State-Of-the-Art Large Language Models}.
\newblock \bibinfo{journal}{\emph{arXiv preprint arXiv:2406.02061}}
  (\bibinfo{year}{2024}).
\newblock


\bibitem[Pedregosa et~al\mbox{.}(2011)]%
        {pedregosa2011scikit}
\bibfield{author}{\bibinfo{person}{Fabian Pedregosa}, \bibinfo{person}{Ga{\"e}l
  Varoquaux}, \bibinfo{person}{Alexandre Gramfort}, \bibinfo{person}{Vincent
  Michel}, \bibinfo{person}{Bertrand Thirion}, \bibinfo{person}{Olivier
  Grisel}, \bibinfo{person}{Mathieu Blondel}, \bibinfo{person}{Peter
  Prettenhofer}, \bibinfo{person}{Ron Weiss}, \bibinfo{person}{Vincent
  Dubourg}, {et~al\mbox{.}}} \bibinfo{year}{2011}\natexlab{}.
\newblock \showarticletitle{Scikit-learn: Machine learning in Python}.
\newblock \bibinfo{journal}{\emph{the Journal of machine Learning research}}
  \bibinfo{volume}{12} (\bibinfo{year}{2011}), \bibinfo{pages}{2825--2830}.
\newblock


\bibitem[Puri et~al\mbox{.}(2021)]%
        {puri_codenet_2021}
\bibfield{author}{\bibinfo{person}{Ruchir Puri}, \bibinfo{person}{David~S.
  Kung}, \bibinfo{person}{Geert Janssen}, \bibinfo{person}{Wei Zhang},
  \bibinfo{person}{Giacomo Domeniconi}, \bibinfo{person}{Vladimir Zolotov},
  \bibinfo{person}{Julian Dolby}, \bibinfo{person}{Jie Chen},
  \bibinfo{person}{Mihir Choudhury}, \bibinfo{person}{Lindsey Decker},
  \bibinfo{person}{Veronika Thost}, \bibinfo{person}{Luca Buratti},
  \bibinfo{person}{Saurabh Pujar}, \bibinfo{person}{Shyam Ramji},
  \bibinfo{person}{Ulrich Finkler}, \bibinfo{person}{Susan Malaika}, {and}
  \bibinfo{person}{Frederick Reiss}.} \bibinfo{year}{2021}\natexlab{}.
\newblock \bibinfo{title}{{CodeNet}: {A} {Large}-{Scale} {AI} for {Code}
  {Dataset} for {Learning} a {Diversity} of {Coding} {Tasks}}.
\newblock
\urldef\tempurl%
\url{http://arxiv.org/abs/2105.12655}
\showURL{%
\tempurl}


\bibitem[Roy and Cordy(2008)]%
        {roy_nicad_2008}
\bibfield{author}{\bibinfo{person}{C.K. Roy} {and} \bibinfo{person}{J.R.
  Cordy}.} \bibinfo{year}{2008}\natexlab{}.
\newblock \showarticletitle{{NICAD}: {Accurate} {Detection} of {Near}-{Miss}
  {Intentional} {Clones} {Using} {Flexible} {Pretty}-{Printing} and {Code}
  {Normalization}}. In \bibinfo{booktitle}{\emph{2008 16th {IEEE}
  {International} {Conference} on {Program} {Comprehension}}}.
  \bibinfo{publisher}{IEEE}, \bibinfo{address}{Amsterdam},
  \bibinfo{pages}{172--181}.
\newblock
\showISBNx{978-0-7695-3176-2}
\href{https://doi.org/10.1109/ICPC.2008.41}{doi:\nolinkurl{10.1109/ICPC.2008.41}}


\bibitem[Roy and Cordy({[n.\,d.]})]%
        {roy_survey_nodate}
\bibfield{author}{\bibinfo{person}{Chanchal~Kumar Roy} {and}
  \bibinfo{person}{James~R Cordy}.} \bibinfo{year}{[n.\,d.]}\natexlab{}.
\newblock \showarticletitle{A {Survey} on {Software} {Clone} {Detection}
  {Research}}.
\newblock  (\bibinfo{year}{[n.\,d.]}).
\newblock


\bibitem[Roy et~al\mbox{.}(2009)]%
        {roy2009comparison}
\bibfield{author}{\bibinfo{person}{Chanchal~K Roy}, \bibinfo{person}{James~R
  Cordy}, {and} \bibinfo{person}{Rainer Koschke}.}
  \bibinfo{year}{2009}\natexlab{}.
\newblock \showarticletitle{Comparison and evaluation of code clone detection
  techniques and tools: A qualitative approach}.
\newblock \bibinfo{journal}{\emph{Science of computer programming}}
  \bibinfo{volume}{74}, \bibinfo{number}{7} (\bibinfo{year}{2009}),
  \bibinfo{pages}{470--495}.
\newblock


\bibitem[Sahoo et~al\mbox{.}(2024)]%
        {sahoo2024systematic}
\bibfield{author}{\bibinfo{person}{Pranab Sahoo}, \bibinfo{person}{Ayush~Kumar
  Singh}, \bibinfo{person}{Sriparna Saha}, \bibinfo{person}{Vinija Jain},
  \bibinfo{person}{Samrat Mondal}, {and} \bibinfo{person}{Aman Chadha}.}
  \bibinfo{year}{2024}\natexlab{}.
\newblock \showarticletitle{A Systematic Survey of Prompt Engineering in Large
  Language Models: Techniques and Applications}.
\newblock \bibinfo{journal}{\emph{arXiv preprint arXiv:2402.07927}}
  (\bibinfo{year}{2024}).
\newblock


\bibitem[Saini et~al\mbox{.}(2018)]%
        {saini_oreo_2018}
\bibfield{author}{\bibinfo{person}{Vaibhav Saini}, \bibinfo{person}{Farima
  Farmahinifarahani}, \bibinfo{person}{Yadong Lu}, \bibinfo{person}{Pierre
  Baldi}, {and} \bibinfo{person}{Cristina~V. Lopes}.}
  \bibinfo{year}{2018}\natexlab{}.
\newblock \showarticletitle{Oreo: detection of clones in the twilight zone}. In
  \bibinfo{booktitle}{\emph{Proceedings of the 2018 26th {ACM} {Joint}
  {Meeting} on {European} {Software} {Engineering} {Conference} and {Symposium}
  on the {Foundations} of {Software} {Engineering}}}. \bibinfo{publisher}{ACM},
  \bibinfo{pages}{354--365}.
\newblock
\showISBNx{978-1-4503-5573-5}
\href{https://doi.org/10.1145/3236024.3236026}{doi:\nolinkurl{10.1145/3236024.3236026}}


\bibitem[Sajnani et~al\mbox{.}(2016)]%
        {sajnani_sourcerercc_2016}
\bibfield{author}{\bibinfo{person}{Hitesh Sajnani}, \bibinfo{person}{Vaibhav
  Saini}, \bibinfo{person}{Jeffrey Svajlenko}, \bibinfo{person}{Chanchal~K.
  Roy}, {and} \bibinfo{person}{Cristina~V. Lopes}.}
  \bibinfo{year}{2016}\natexlab{}.
\newblock \showarticletitle{{SourcererCC}: scaling code clone detection to
  big-code}. In \bibinfo{booktitle}{\emph{Proceedings of the 38th
  {International} {Conference} on {Software} {Engineering}}}.
  \bibinfo{publisher}{ACM}, \bibinfo{address}{Austin Texas},
  \bibinfo{pages}{1157--1168}.
\newblock
\showISBNx{978-1-4503-3900-1}
\href{https://doi.org/10.1145/2884781.2884877}{doi:\nolinkurl{10.1145/2884781.2884877}}


\bibitem[Sorokin et~al\mbox{.}(2023)]%
        {sorokin_cct-code_2023}
\bibfield{author}{\bibinfo{person}{Nikita Sorokin}, \bibinfo{person}{Dmitry
  Abulkhanov}, \bibinfo{person}{Sergey Nikolenko}, {and}
  \bibinfo{person}{Valentin Malykh}.} \bibinfo{year}{2023}\natexlab{}.
\newblock \bibinfo{title}{{CCT}-{Code}: {Cross}-{Consistency} {Training} for
  {Multilingual} {Clone} {Detection} and {Code} {Search}}.
\newblock
\urldef\tempurl%
\url{http://arxiv.org/abs/2305.11626}
\showURL{%
\tempurl}


\bibitem[Tao et~al\mbox{.}(2022)]%
        {tao_c4_2022}
\bibfield{author}{\bibinfo{person}{Chenning Tao}, \bibinfo{person}{Qi Zhan},
  \bibinfo{person}{Xing Hu}, {and} \bibinfo{person}{Xin Xia}.}
  \bibinfo{year}{2022}\natexlab{}.
\newblock \showarticletitle{C4: contrastive cross-language code clone
  detection}. In \bibinfo{booktitle}{\emph{Proceedings of the 30th {IEEE}/{ACM}
  {International} {Conference} on {Program} {Comprehension}}}.
  \bibinfo{publisher}{ACM}, \bibinfo{address}{Virtual Event},
  \bibinfo{pages}{413--424}.
\newblock
\showISBNx{978-1-4503-9298-3}
\href{https://doi.org/10.1145/3524610.3527911}{doi:\nolinkurl{10.1145/3524610.3527911}}


\bibitem[Touvron et~al\mbox{.}(2023)]%
        {touvron_llama_2023}
\bibfield{author}{\bibinfo{person}{Hugo Touvron}, \bibinfo{person}{Louis
  Martin}, \bibinfo{person}{Kevin Stone}, \bibinfo{person}{Peter Albert},
  \bibinfo{person}{Amjad Almahairi}, \bibinfo{person}{Yasmine Babaei},
  \bibinfo{person}{Nikolay Bashlykov}, \bibinfo{person}{Soumya Batra},
  \bibinfo{person}{Prajjwal Bhargava}, \bibinfo{person}{Shruti Bhosale},
  \bibinfo{person}{Dan Bikel}, \bibinfo{person}{Lukas Blecher},
  \bibinfo{person}{Cristian~Canton Ferrer}, \bibinfo{person}{Moya Chen},
  \bibinfo{person}{Guillem Cucurull}, \bibinfo{person}{David Esiobu},
  \bibinfo{person}{Jude Fernandes}, \bibinfo{person}{Jeremy Fu},
  \bibinfo{person}{Wenyin Fu}, \bibinfo{person}{Brian Fuller},
  \bibinfo{person}{Cynthia Gao}, \bibinfo{person}{Vedanuj Goswami},
  \bibinfo{person}{Naman Goyal}, \bibinfo{person}{Anthony Hartshorn},
  \bibinfo{person}{Saghar Hosseini}, \bibinfo{person}{Rui Hou},
  \bibinfo{person}{Hakan Inan}, \bibinfo{person}{Marcin Kardas},
  \bibinfo{person}{Viktor Kerkez}, \bibinfo{person}{Madian Khabsa},
  \bibinfo{person}{Isabel Kloumann}, \bibinfo{person}{Artem Korenev},
  \bibinfo{person}{Punit~Singh Koura}, \bibinfo{person}{Marie-Anne Lachaux},
  \bibinfo{person}{Thibaut Lavril}, \bibinfo{person}{Jenya Lee},
  \bibinfo{person}{Diana Liskovich}, \bibinfo{person}{Yinghai Lu},
  \bibinfo{person}{Yuning Mao}, \bibinfo{person}{Xavier Martinet},
  \bibinfo{person}{Todor Mihaylov}, \bibinfo{person}{Pushkar Mishra},
  \bibinfo{person}{Igor Molybog}, \bibinfo{person}{Yixin Nie},
  \bibinfo{person}{Andrew Poulton}, \bibinfo{person}{Jeremy Reizenstein},
  \bibinfo{person}{Rashi Rungta}, \bibinfo{person}{Kalyan Saladi},
  \bibinfo{person}{Alan Schelten}, \bibinfo{person}{Ruan Silva},
  \bibinfo{person}{Eric~Michael Smith}, \bibinfo{person}{Ranjan Subramanian},
  \bibinfo{person}{Xiaoqing~Ellen Tan}, \bibinfo{person}{Binh Tang},
  \bibinfo{person}{Ross Taylor}, \bibinfo{person}{Adina Williams},
  \bibinfo{person}{Jian~Xiang Kuan}, \bibinfo{person}{Puxin Xu},
  \bibinfo{person}{Zheng Yan}, \bibinfo{person}{Iliyan Zarov},
  \bibinfo{person}{Yuchen Zhang}, \bibinfo{person}{Angela Fan},
  \bibinfo{person}{Melanie Kambadur}, \bibinfo{person}{Sharan Narang},
  \bibinfo{person}{Aurelien Rodriguez}, \bibinfo{person}{Robert Stojnic},
  \bibinfo{person}{Sergey Edunov}, {and} \bibinfo{person}{Thomas Scialom}.}
  \bibinfo{year}{2023}\natexlab{}.
\newblock \bibinfo{title}{Llama 2: {Open} {Foundation} and {Fine}-{Tuned}
  {Chat} {Models}}.
\newblock
\urldef\tempurl%
\url{http://arxiv.org/abs/2307.09288}
\showURL{%
\tempurl}


\bibitem[Tunstall et~al\mbox{.}(2023)]%
        {Tunstall2023starchat-alpha}
\bibfield{author}{\bibinfo{person}{Lewis Tunstall}, \bibinfo{person}{Nathan
  Lambert}, \bibinfo{person}{Nazneen Rajani}, \bibinfo{person}{Edward
  Beeching}, \bibinfo{person}{Teven Le~Scao}, \bibinfo{person}{Leandro von
  Werra}, \bibinfo{person}{Sheon Han}, \bibinfo{person}{Philipp Schmid}, {and}
  \bibinfo{person}{Alexander Rush}.} \bibinfo{year}{2023}\natexlab{}.
\newblock \showarticletitle{Creating a Coding Assistant with StarCoder}.
\newblock \bibinfo{journal}{\emph{Hugging Face Blog}} (\bibinfo{year}{2023}).
\newblock


\bibitem[Vislavski et~al\mbox{.}(2018)]%
        {vislavski_licca_2018}
\bibfield{author}{\bibinfo{person}{Tijana Vislavski}, \bibinfo{person}{Gordana
  Rakic}, \bibinfo{person}{Nicolas Cardozo}, {and} \bibinfo{person}{Zoran
  Budimac}.} \bibinfo{year}{2018}\natexlab{}.
\newblock \showarticletitle{{LICCA}: {A} tool for cross-language clone
  detection}. In \bibinfo{booktitle}{\emph{2018 {IEEE} 25th {International}
  {Conference} on {Software} {Analysis}, {Evolution} and {Reengineering}
  ({SANER})}}. \bibinfo{publisher}{IEEE}, \bibinfo{address}{Campobasso},
  \bibinfo{pages}{512--516}.
\newblock
\showISBNx{978-1-5386-4969-5}
\href{https://doi.org/10.1109/SANER.2018.8330250}{doi:\nolinkurl{10.1109/SANER.2018.8330250}}


\bibitem[Wang et~al\mbox{.}(2020)]%
        {wang_detecting_2020}
\bibfield{author}{\bibinfo{person}{Wenhan Wang}, \bibinfo{person}{Ge Li},
  \bibinfo{person}{Bo Ma}, \bibinfo{person}{Xin Xia}, {and}
  \bibinfo{person}{Zhi Jin}.} \bibinfo{year}{2020}\natexlab{}.
\newblock \showarticletitle{Detecting {Code} {Clones} with {Graph} {Neural}
  {Network} and {Flow}-{Augmented} {Abstract} {Syntax} {Tree}}. In
  \bibinfo{booktitle}{\emph{2020 {IEEE} 27th {International} {Conference} on
  {Software} {Analysis}, {Evolution} and {Reengineering} ({SANER})}}.
  \bibinfo{publisher}{IEEE}, \bibinfo{address}{London, ON, Canada},
  \bibinfo{pages}{261--271}.
\newblock
\showISBNx{978-1-72815-143-4}
\href{https://doi.org/10.1109/SANER48275.2020.9054857}{doi:\nolinkurl{10.1109/SANER48275.2020.9054857}}


\bibitem[Wei and Li(2017)]%
        {wei_supervised_2017}
\bibfield{author}{\bibinfo{person}{Huihui Wei} {and} \bibinfo{person}{Ming
  Li}.} \bibinfo{year}{2017}\natexlab{}.
\newblock \showarticletitle{Supervised {Deep} {Features} for {Software}
  {Functional} {Clone} {Detection} by {Exploiting} {Lexical} and {Syntactical}
  {Information} in {Source} {Code}}. In \bibinfo{booktitle}{\emph{Proceedings
  of the {Twenty}-{Sixth} {International} {Joint} {Conference} on {Artificial}
  {Intelligence}}}. \bibinfo{publisher}{International Joint Conferences on
  Artificial Intelligence Organization}, \bibinfo{address}{Melbourne,
  Australia}, \bibinfo{pages}{3034--3040}.
\newblock
\showISBNx{978-0-9992411-0-3}
\href{https://doi.org/10.24963/ijcai.2017/423}{doi:\nolinkurl{10.24963/ijcai.2017/423}}


\bibitem[Wei et~al\mbox{.}(2024)]%
        {StarCoder2-Instruct}
\bibfield{author}{\bibinfo{person}{Yuxiang Wei}, \bibinfo{person}{Federico
  Cassano}, \bibinfo{person}{Jiawei Liu}, \bibinfo{person}{Yifeng Ding},
  \bibinfo{person}{Naman Jain}, \bibinfo{person}{Harm de Vries},
  \bibinfo{person}{Leandro von Werra}, \bibinfo{person}{Arjun Guha}, {and}
  \bibinfo{person}{Lingming Zhang}.} \bibinfo{year}{2024}\natexlab{}.
\newblock \bibinfo{title}{StarCoder2-Instruct: Fully Transparent and Permissive
  Self-Alignment for Code Generation}.
\newblock
\urldef\tempurl%
\url{https://github.com/bigcode-project/starcoder2-self-align}
\showURL{%
\tempurl}


\bibitem[White et~al\mbox{.}(2016)]%
        {white_deep_2016}
\bibfield{author}{\bibinfo{person}{Martin White}, \bibinfo{person}{Michele
  Tufano}, \bibinfo{person}{Christopher Vendome}, {and} \bibinfo{person}{Denys
  Poshyvanyk}.} \bibinfo{year}{2016}\natexlab{}.
\newblock \showarticletitle{Deep learning code fragments for code clone
  detection}. In \bibinfo{booktitle}{\emph{Proceedings of the 31st {IEEE}/{ACM}
  {International} {Conference} on {Automated} {Software} {Engineering}}}.
  \bibinfo{publisher}{ACM}, \bibinfo{address}{Singapore Singapore},
  \bibinfo{pages}{87--98}.
\newblock
\showISBNx{978-1-4503-3845-5}
\href{https://doi.org/10.1145/2970276.2970326}{doi:\nolinkurl{10.1145/2970276.2970326}}


\bibitem[Ye et~al\mbox{.}(2024)]%
        {ye_prompt_2024}
\bibfield{author}{\bibinfo{person}{Qinyuan Ye}, \bibinfo{person}{Maxamed
  Axmed}, \bibinfo{person}{Reid Pryzant}, {and} \bibinfo{person}{Fereshte
  Khani}.} \bibinfo{year}{2024}\natexlab{}.
\newblock \bibinfo{title}{Prompt {Engineering} a {Prompt} {Engineer}}.
\newblock
\urldef\tempurl%
\url{http://arxiv.org/abs/2311.05661}
\showURL{%
\tempurl}


\bibitem[Yin et~al\mbox{.}(2024)]%
        {yin2024should}
\bibfield{author}{\bibinfo{person}{Ziqi Yin}, \bibinfo{person}{Hao Wang},
  \bibinfo{person}{Kaito Horio}, \bibinfo{person}{Daisuke Kawahara}, {and}
  \bibinfo{person}{Satoshi Sekine}.} \bibinfo{year}{2024}\natexlab{}.
\newblock \showarticletitle{Should We Respect LLMs? A Cross-Lingual Study on
  the Influence of Prompt Politeness on LLM Performance}.
\newblock \bibinfo{journal}{\emph{arXiv preprint arXiv:2402.14531}}
  (\bibinfo{year}{2024}).
\newblock


\bibitem[Zhao and Huang(2018)]%
        {zhao_deepsim_2018}
\bibfield{author}{\bibinfo{person}{Gang Zhao} {and} \bibinfo{person}{Jeff
  Huang}.} \bibinfo{year}{2018}\natexlab{}.
\newblock \showarticletitle{{DeepSim}: deep learning code functional
  similarity}. In \bibinfo{booktitle}{\emph{Proceedings of the 2018 26th {ACM}
  {Joint} {Meeting} on {European} {Software} {Engineering} {Conference} and
  {Symposium} on the {Foundations} of {Software} {Engineering}}}.
  \bibinfo{publisher}{ACM}, \bibinfo{pages}{141--151}.
\newblock
\showISBNx{978-1-4503-5573-5}
\href{https://doi.org/10.1145/3236024.3236068}{doi:\nolinkurl{10.1145/3236024.3236068}}


\bibitem[Zhu et~al\mbox{.}(2022)]%
        {zhu_xlcost_2022}
\bibfield{author}{\bibinfo{person}{Ming Zhu}, \bibinfo{person}{Aneesh Jain},
  \bibinfo{person}{Karthik Suresh}, \bibinfo{person}{Roshan Ravindran},
  \bibinfo{person}{Sindhu Tipirneni}, {and} \bibinfo{person}{Chandan~K.
  Reddy}.} \bibinfo{year}{2022}\natexlab{}.
\newblock \bibinfo{title}{{XLCoST}: {A} {Benchmark} {Dataset} for
  {Cross}-lingual {Code} {Intelligence}}.
\newblock
\urldef\tempurl%
\url{http://arxiv.org/abs/2206.08474}
\showURL{%
\tempurl}


\end{thebibliography}

\end{document}